\documentclass[conference]{IEEEtran}
\IEEEoverridecommandlockouts
\usepackage{cite}
\usepackage{amsmath,amssymb,amsfonts}
\usepackage{algorithmic}
\usepackage{graphicx}
\usepackage{textcomp}
\usepackage{xcolor}
\usepackage[hyphens]{url}

\usepackage{mathptmx}
\usepackage{fancyhdr}
\usepackage{soul}
\usepackage[normalem]{ulem}
\usepackage{microtype}
\usepackage{booktabs}
\usepackage{multirow}
\usepackage{algorithm}
\usepackage{xspace}
\usepackage[font=small,labelfont=bf]{caption}
\usepackage[caption=false]{subfig}
\usepackage{mathtools}
\usepackage{pifont}
\usepackage{color}
\usepackage{makecell}
\usepackage{lipsum}
\usepackage{subfig}
\usepackage{tikz}
\usepackage{enumitem}
\usepackage{hyperref}

\def\BibTeX{{\rm B\kern-.05em{\sc i\kern-.025em b}\kern-.08em
    T\kern-.1667em\lower.7ex\hbox{E}\kern-.125emX}}
\begin{document}

%%%%%%%%%%%%%%%%%%%%%%%%
\newcommand{\todo}[1]{\textcolor{red}{#1}\xspace}
\newcommand{\remove}[1]{\textcolor{blue}{\sout{#1}}\xspace}
\newcommand{\new}[1]{\textcolor{blue}{#1}\xspace}
\newcommand{\move}[1]{\textcolor{green}{#1}\xspace}

\newcommand{\old}[1]{}
\newcommand{\fig}[1]{Figure~\ref{#1}}
\newcommand{\sect}[1]{Section~\ref{#1}}
\newcommand{\tab}[1]{Table~\ref{#1}}
\newcommand{\algo}[1]{Algorithm~\ref{#1}}
\newcommand{\eqn}[1]{Equation~\ref{#1}}
\newcommand{\e}[1]{\times 10^{#1}}

\newcommand{\dmaload}[0]{\texttt{LOAD\_TILE}\xspace}
\newcommand{\dmastore}[0]{\texttt{STORE\_TILE}\xspace}
\newcommand{\gemm}[0]{\texttt{GEMM\_OP}\xspace}
\newcommand{\conv}[0]{\texttt{CONV\_OP}\xspace}
\newcommand{\vect}[0]{\texttt{VECTOR\_OP}\xspace}
\newcommand{\pluseq}{\mathrel{+}=}

\newcommand{\baseline}[0]{$Disagg$\xspace}
\newcommand{\ours}[0]{$PreSto$\xspace}

\title{\huge PreSto: An In-Storage Data Preprocessing System\\ for Training Recommendation Models
\thanks{
\IEEEauthorrefmark{2} Co-first authors who contributed equally to this research.\newline
\noindent\rule{4cm}{0.4pt}\newline
This is the author preprint version of the work. The authoritative version will appear in the Proceedings of the 51st IEEE/ACM International Symposium on Computer Architecture (ISCA-51), 2024.
}
}

\author{
  \IEEEauthorblockN{Yunjae Lee\IEEEauthorrefmark{2}}
  \IEEEauthorblockA{School of Electrical Engineering\\
  KAIST\\
  \textit{yunjae408@kaist.ac.kr}}
  \and
  \IEEEauthorblockN{Hyeseong Kim\IEEEauthorrefmark{2}}
  \IEEEauthorblockA{School of Electrical Engineering\\
  KAIST\\
  \textit{hyeseong.kim@kaist.ac.kr}}
  \and
  \IEEEauthorblockN{Minsoo Rhu}
  \IEEEauthorblockA{School of Electrical Engineering\\
  KAIST\\
  \textit{mrhu@kaist.ac.kr}}
}

\maketitle

\begin{abstract}

Training recommendation systems (RecSys) faces
several challenges as it	requires the ``data preprocessing'' stage to preprocess
an ample amount of raw data and feed them to the GPU for training in a
seamless manner.  To sustain  high training throughput, state-of-the-art
solutions reserve a large fleet of CPU servers for preprocessing
which  incurs substantial deployment cost and power consumption.
Our characterization reveals that prior CPU-centric preprocessing
 is bottlenecked on feature generation and feature normalization
operations as it fails to reap out the abundant inter-/intra-feature
parallelism in RecSys preprocessing.  PreSto
 is a storage-centric preprocessing system leveraging In-Storage
Processing (ISP), which offloads the bottlenecked preprocessing
operations to our ISP units.  We show that PreSto outperforms the baseline
CPU-centric system with a $9.6\times$ speedup in end-to-end
preprocessing time, $4.3\times$ enhancement in cost-efficiency, and 
$11.3\times$ improvement in energy-efficiency on average for
production-scale RecSys preprocessing.
  
  \end{abstract}

\hypersetup{hidelinks}


\section{Introduction}
\label{sect:introduction}

	Deep neural network (DNN) based machine learning (ML) algorithms have
demonstrated their effectiveness in a wide range of application domains.
Among the successfully deployed ML applications, recommendation systems
(RecSys) have emerged as a highly effective tool for online content
recommendation services.
Such rising demand for recommendation services has rendered hyperscalers to
dedicate significant resources to the development and training of diverse
RecSys models to ensure high-quality inference services. Unlike
latency-optimized ML inference, training algorithms are
throughput-hungry workloads that favor high-performance, throughput-optimized
accelerators like GPUs.  However, these power-hungry GPUs 
account for a large portion of ML system's operating expenses, so
maintaining high GPU utilization becomes critical for lowering
 TCO (total cost of ownership).  Unfortunately,
						keeping the RecSys training pipeline busy with minimal GPU idle
						time requires the ``data preprocessing'' stage to 
						preprocess an ample amount of raw data, so that
						the preprocessed, train-ready tensors can be fed into the GPU in a
						seamless manner.

Traditionally, the RecSys training pipeline employed \emph{offline} data preprocessing
where the raw data retrieved from the storage system is transformed into
train-ready tensors in advance and gets archived at a separate storage space for future
usage.  However, the proliferation of petabyte-scale
data and the wide variety of RecSys models developed by ML engineers have
rendered offline preprocessing to incur an intractable amount of
overhead~\cite{dsi}. Specifically,  it becomes increasingly difficult to provision the
substantial storage space required to store all the data preprocessed offline, while
also adapting to changes in the newly developed RecSys models. 

These challenges have triggered a shift towards \emph{online} preprocessing,
			which involves preprocessing the raw data ``on-the-fly''. Online
			preprocessing obviates the need to separately store the preprocessed
			data, so it helps better respond to changes in the model
			architecture~\cite{dsi}.  Nevertheless, these online preprocessing
			approaches introduce several system-level challenges, potentially causing
			a throughput mismatch between data preprocessing and ML model training.
			Consider a scenario where preprocessing and training jobs are
			\emph{co-located} within the same GPU training server node, i.e.,
			preprocessing is undertaken on the host CPU that also manages the
			GPU-side training job. If the CPU does not have a high enough computation
			power for preprocessing, it fails in generating sufficient amount of
			train-ready tensors that the GPU can consume, leading to significant GPU
			underutilization (less than $20\%$ GPU utility, \sect{sect:motivation}).
			To address these challenges, Zhao et al.~\cite{dsi} and Audibert et
			al.~\cite{disaggregation_google} suggest a \emph{server disaggregation} solution where
			a pool of CPU servers is reserved for
			preprocessing purposes.  Disaggregating CPU servers for preprocessing
			allows hundreds to thousands of CPU cores to be allocated
			\emph{on-demand}, even for a single data preprocessing job, effectively
			closing the performance gap between preprocessing and model training
			thereby minimizing GPU idle time~\cite{dsi,xdl}. However, this baseline ``CPU-centric'' disaggregated preprocessing 
			incurs significant deployment cost and power consumption due to
			the large number of pooled CPU servers.

Given this landscape, an important objective of our work is to characterize
baseline CPU-centric disaggregated data preprocessing systems targeting
production-scale RecSys models, root-causing its critical system-level
challenges. A key observation we make is that the majority of data
preprocessing time is spent conducting \emph{feature generation} and
\emph{feature normalization} operations, which inherently contain high
\emph{inter-/intra-feature parallelism}. However, the latency-optimized CPU
architectures, the de facto standard in data preprocessing, fail to fully
exploit the abundant inter-/intra-feature parallelism in feature
generation and normalization, leading to sub-optimal performance. This in turn
leads to the feature generation and normalization to account for $79\%$
of the RecSys data preprocessing time, causing the most significant performance
bottleneck. To make up for the meager preprocessing throughput provided with
CPUs, the data preprocessing stage necessitates a large number of CPU cores (up
		to several hundreds) to be allocated  so that its aggregate preprocessing
throughput matches the throughput demands of GPU's model training stage, which
leads to substantial deployment cost.

In this work, we propose to employ \emph{accelerated} computing for RecSys data
preprocessing to fundamentally address its system-level challenges at low cost.
Because the abundant inter-/intra-feature parallelism in data preprocessing is
well-suited for domain-specific acceleration, our first key proposal is to
\emph{offload} the time-consuming feature generation and normalization
operations to our accelerator for high-performance data preprocessing.  An
important design decision still remains, however, regarding \emph{where} our
data preprocessing accelerator should be placed within the overall system
architecture. State-of-the-art data preprocessing solutions
employ disaggregated CPU servers to dynamically allocate the right amount of
CPU cores  for data preprocessing~\cite{dsi,disaggregation_google}.  While using our data
preprocessing accelerator as a drop-in replacement for CPUs within the
disaggregated CPU servers is a practically feasible option, we observe that
such a design point is sub-optimal as it unnecessarily incurs high network
traffic to copy data in (the raw data to be preprocessed) and out (the
		preprocessed train-ready tensors) of the disaggregated server for data
preprocessing.

To this end, we present \ours ({\bf Pre}processing in-{\bf Sto}rage), which is
an In-Storage Processing (ISP) based data preprocessing system for RecSys
training. In conventional systems, the petabyte-scale raw data that is to be preprocessed are stored in a distributed storage system.
Rather than copying the raw data over the datacenter network and preprocessing
them at a disaggregated, \emph{remote} preprocessing server, \ours conducts
preprocessing \emph{near data} using ISP. We demonstrate that such
``storage-centric'' data preprocessing can effectively close
the performance gap between preprocessing and model training at a much lower
cost compared to existing solutions. 
Overall, \ours provides
high speedup on data preprocessing performance (average
$9.6\times$) at low cost, significantly reducing the TCO (average $4.3\times$) and energy consumption (average $11.3\times$) 
vs. the baseline CPU-centric disaggregated preprocessing.

\section{Background}
\label{sect:background}

\subsection{End-to-End RecSys Training Pipeline}
\label{sect:recsys_training_pipeline}

DNNs typically require some form of input data preprocessing before
training. For instance, image classification requires
preprocessing operations like image decoding, resizing, cropping, or
augmentation. Additionally, speech recognition preprocessing includes Fourier
transform and normalization of audio data~\cite{trainbox}. Similarly, RecSys also requires 
a unique data preprocessing to generate the train-ready tensors consumed
by the  model training stage, which we detail below.

Traditionally, the RecSys training pipeline employed \emph{offline} data
preprocessing where the raw feature data stored in the storage system is
transformed into train-ready tensors well before model training takes
place. However, the proliferation of petabyte-scale data and the frequent
development of diverse RecSys models by ML engineers have made offline
preprocessing impractical because it is difficult to manage the substantial
storage space required to store the offline preprocessed data. This challenge
has triggered a shift towards \emph{online} preprocessing~\cite{dsi}, which involves
preprocessing the raw feature data ``on-the-fly'' (\fig{fig:recsys_pipeline}). The train-ready tensors are derived from the 
features that are constantly generated online by inference services.
Specifically, inference servers log various end-user's interactions with the
inference service as distinct features (e.g., news feed a user has clicked,
		items a user has purchased) using logging engines, e.g., Meta's Scribe\cite{scribe}. 
		Additionally, various streaming and batch engines, such as Spark\cite{spark}, further label and filter data before storing them 
		in a centralized data warehouse\cite{tectonic_shift}. These raw feature data are categorized into two types: dense and sparse. Dense features represent continuous values
(e.g., the time when a user viewed a video from YouTube), while sparse features
represent sparse categorical values which can be variable-length (e.g., list of
		YouTube videos a user has viewed over a one-hour period). The logged features archived within the centralized data warehouse are then fetched into the storage system of each datacenter for future data preprocessing. 
		The data preprocessing stage, also known as the ETL ({\bf E}xtract, {\bf T}ransform, and
		{\bf L}oad) phase, involves the following series of operations:

\begin{figure*}[t!]
\centering
\includegraphics[width=.96\textwidth]{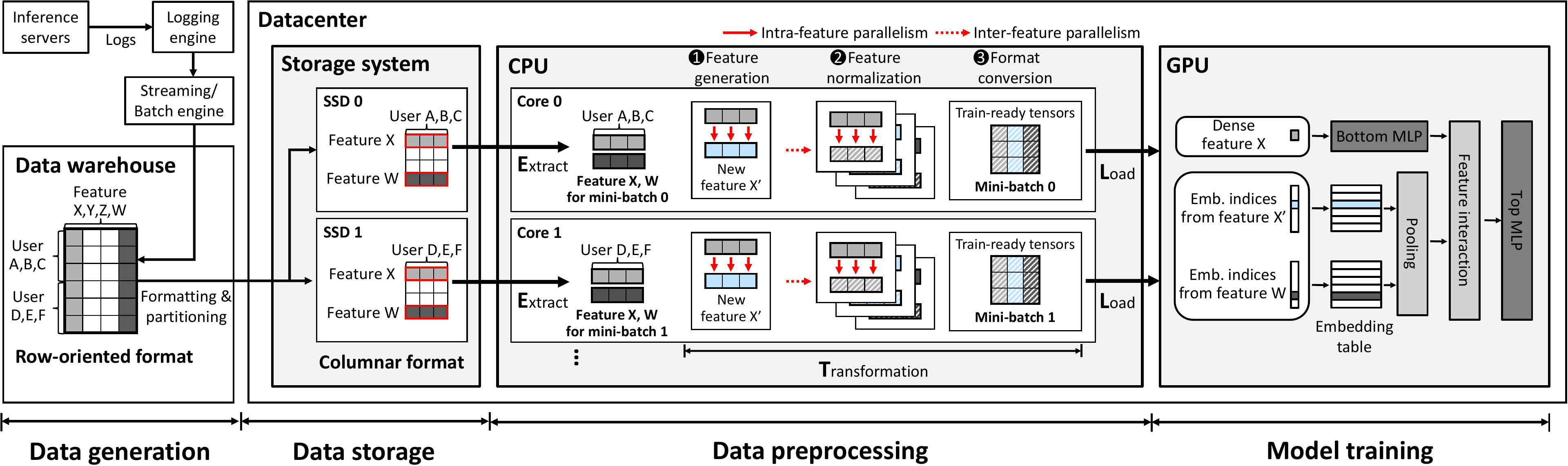} 
\caption{High-level overview of the end-to-end RecSys training pipeline. In this work, we assume our baseline data storage and ingestion pipeline for data preprocessing by referring to the related academic literature published by Meta~\cite{dsi,scribe,tectonic_shift,recd}.}
\label{fig:recsys_pipeline}
\vspace{-1.3em}
\end{figure*}

\begin{itemize}
\item (Extract) The logged raw feature data
are first retrieved from the storage system in preparation for the feature-specific
transform operations.

\item (Transform) The extracted raw feature data go through feature generation and feature
normalization in order to generate the train-ready tensors. 

\item (Load) The train-ready tensors are copied over to the GPU's high-bandwidth memory (HBM) in preparation for the RecSys model training.

\end{itemize}

Once the train-ready tensors are loaded into GPU's memory, the actual
model training is undertaken. Specifically, the GPU executes the
embedding layers (embedding lookups, pooling for embedding reductions), feature interactions (batched GEMM), and MLP layers (GEMM).
When it comes to embedding layers,
			 the embedding look-up operation retrieves embeddings
			 from an embedding table~\cite{facebook_dlrm}, which utilizes embedding
			 indices that are generated during the Transform phase of ETL (e.g., the
					 embedding indices transformed from feature $X$' and $W$ in \fig{fig:recsys_pipeline} are utilized for
					 embedding look-up operations). Recent work on various system-level performance optimizations
			 for RecSys has primarily focused on this ``training'' stage
			 of end-to-end training pipeline, rather than data preprocessing. The focus
			 of this work is on RecSys data preprocessing, so we refer
			 to these relevant prior studies for more details on the RecSys model architectures and 
			 training/inference~\cite{tensordimm,centaur,facebook_dlrm,udit:resys:hpca2020:industry_track,acun:resys:hpca2021:industry_track,isca2022_mudigere,fafnir,recnmp,tensorcasting,trim,tcaching,merci,recssd,rm_ssd,deeprecsys,hercules,mprec,recpipe,facebook_hpca2018}.

While GPU-centric systems are popular options for training purposes, it is
worth emphasizing that the entire data preprocessing stage (the ETL phase) is
executed using CPUs in state-of-the-art data preprocessing
systems~\cite{tf_data,vldb2021_mohan,cachew,fastflow,disaggregation_google,plumber_mlsys2022},
	including those for RecSys~\cite{dsi,intune_recsys2023,recd,xdl}. In the rest of
	this paper, we assume such ``CPU-centric'' RecSys data preprocessing system
	for our baseline system.

\subsection{Key Properties of RecSys Data Preprocessing}
\label{sect:key_property_recsys_preproc}

While numerous prior work explored preprocessing for vision, speech,
			and language processing~\cite{vldb2021_mohan, tf_data, nvidia_dali, disaggregation_google, 
			plumber_mlsys2022, cachew, fastflow, quiver, dlbooster,icache}, data preprocessing for RecSys
				is relatively less explored~\cite{dsi,recd,intune_recsys2023,xdl}. Unlike image or audio
				data, RecSys data is represented in a \emph{tabular} format with
				multiple rows and columns. Concretely, each row represents an
				individual ``user'' whereas each column represents a distinct
				``feature'' related to that user's past interactions with the RecSys
				inference service (shown as the data generation stage in
						\fig{fig:recsys_pipeline}). 

In the context of online preprocessing, deciding which features
to utilize for model training depends on the ML engineer's choice, i.e., it is extremely challenging to predict which
specific features will be utilized. Consequently, the
hardware/software system for online preprocessing exhibits some unique
properties in the Extract and Transform phase:

\begin{itemize}
\item (Extract) The raw feature data is first converted and stored in a
\emph{columnar} format (shown as the data storage stage in \fig{fig:recsys_pipeline}). A group of rows are sharded into mutually
exclusive \emph{partitions} and different partitions are stored as independent
columnar files into a distributed storage system of datacenter (e.g., the two partitions in
\fig{fig:recsys_pipeline} are stored as two separate columnar files over two
		SSDs).  The reason why these tabular data are converted in a columnar
format is to avoid overfetching unwanted features.  For example, in the data
storage stage depicted in \fig{fig:recsys_pipeline}, with a columnar format,
				any given feature for all the users can be \emph{selectively} extracted
				from the storage system without having to retrieve unwanted features,
				e.g., it is possible to only fetch features $X$ and $W$ without having
				to fetch features $Y$ and $Z$. In contrast, with the original,
				row-oriented format, extracting features $X$ and $W$ for all users
				inevitably leads to (unwanted) features $Y$ and $Z$ to be retrieved,
				wasting data read bandwidth.

\item (Transform)  The transformations conducted at this phase generate the
mini-batch inputs that are utilized by the GPU for model training. All
transformation operations performed within a mini-batch (detailed in the next
		subsection) are executed independently from transformations targeting other
mini-batches. This is because RecSys transformation operations exhibit abundant
\emph{inter-/intra-feature parallelism}.  Specifically, each element within a
given feature vector (e.g., user $A$ and $C$'s feature $X$ in
		\fig{fig:recsys_pipeline}) represents a given user's interaction and there
exists no data dependency across different users.  Therefore, a transformation
operation can be conducted element-wise by exploiting \emph{intra}-feature
parallelism. Similarly, different features (e.g., feature $X$ and $W$ for all
		users) are subject to independent transformation operations by leveraging
\emph{inter}-feature parallelism.

	\end{itemize}

\subsection{Feature Generation/Normalization in Data Preprocessing}
\label{subsect:feature_gen}

RecSys data preprocessing can be divided into
three key steps. First, the \emph{feature generation} step generates new features
using the raw feature data extracted from storage (Step \ding{182} in
		\fig{fig:recsys_pipeline}, e.g., a new feature $X$' is generated from the
		raw feature $X$). Notably, one of the representative sparse feature
generation operations is the ``Bucketize''
operation~\cite{dsi,torcharrow:github}, which transforms dense features into
sparse features by sharding features based on predefined bucket boundaries
(\algo{algo:bucketize}\footnote{This paper focuses on the feature generation
 (Bucketize) and feature normalization (SigridHash, Log) operations publicly available
 in the open-source TorchArrow~\cite{torcharrow:github}. \algo{algo:bucketize}
 and \algo{algo:sigridhash} are simplified versions of each algorithm
 implemented in TorchArrow.  }). Once the desired number of features is
generated, they undergo \emph{feature normalization} (Step \ding{183}). Common
feature normalization techniques include ``Log'' (which normalizes dense
		features using a logarithmic function) and
``SigridHash''~\cite{dsi,torcharrow:github} (which normalizes sparse features
		by computing a hash value that maps those features within the maximum index
		of the corresponding embedding table of the RecSys model, see
		\algo{algo:sigridhash}). Finally, the normalized features are converted
into an input mini-batch (Step \ding{184}), which eventually gets loaded into
GPU memory for training the RecSys model. 

\begin{algorithm}[t]
	\caption{Bucketize for feature generation~\cite{torcharrow:github}}
	\label{algo:bucketize}
	\begin{algorithmic}[1]
		\STATE Input dense feature $a[1\,\ldots\,n]$; bucket boundary $b[1\,\ldots\,m]$; output $c[1\,\ldots\,n]$
		\STATE /* Digitize input dense features based on bucket */
		\FOR {$i \leftarrow$ $1$ to $n$}
		\STATE /* Find the index of buckets to which the input value belongs using binary search algorithm*/
		\STATE $c[i] \leftarrow \texttt{SearchBucketID}(a[i],\,b[1\,\ldots\,m])$
		\ENDFOR
	\end{algorithmic} 
	\end{algorithm}

	\begin{algorithm}[t]
	\caption{SigridHash for feature normalization~\cite{torcharrow:github}} 
	\label{algo:sigridhash}
	\begin{algorithmic}[1]
		\STATE Input sparse feature $a[1\,\ldots\,n]$; seed $s$; max value $d$; output $c[1\,\ldots\,n]$;
		\STATE /* Apply hashing to input sparse features and limit their values */
		\FOR {$i \leftarrow$ $1$ to $n$}
		\STATE /* Compute seeded hash function */
		\STATE $h \leftarrow \texttt{ComputeHash}(a[i],\,s)$
		\STATE $c[i] \leftarrow$ $h\bmod d$
		\ENDFOR
  \end{algorithmic} 
  \end{algorithm}

\subsection{System Architecture for CPU-centric Data Preprocessing}
\label{sect:system_architecture_recsys}

{\bf Software architecture.} As shown in~\fig{fig:recsys_pipeline}, the RecSys training pipeline employs the
producer-consumer model. GPU training workers load and consume train-ready
tensors (i.e., mini-batches) that the CPU preprocessing workers generate by
transforming raw feature data.  Because all transformation operations conducted
within a mini-batch are executed locally without any dependencies to other
mini-batches, state-of-the-art frameworks for end-to-end RecSys training pipeline such as
TorchRec~\cite{torchrec} allocate a single worker per each CPU core to handle the generation of
each train-ready tensors that constitute a given mini-batch. By spawning
multiple CPU workers in parallel, multiple input mini-batches are concurrently
generated, enabling scalable improvements in data preprocessing throughput.

\begin{figure}[t]
	\centering
	\includegraphics[width=0.46\textwidth]{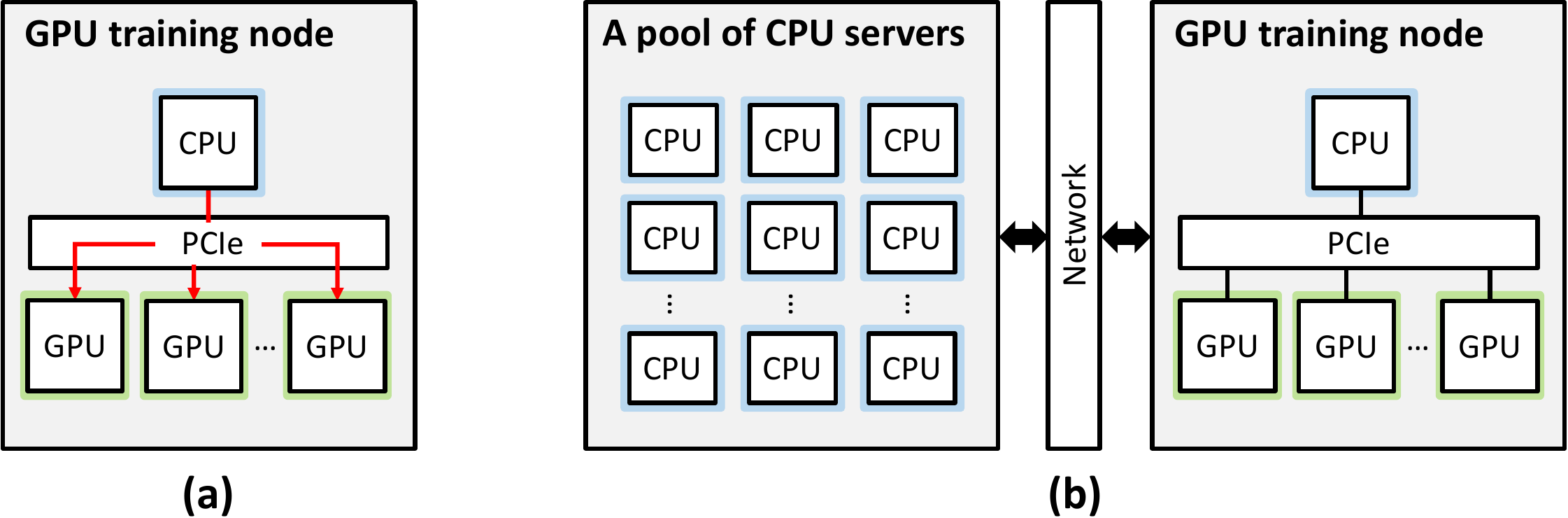} 
	\caption{System architectures for RecSys training.  (a) A system that co-locates CPU-based data preprocessing
		workers with GPU-based model training workers within the same server node. (b) A system that provisions a pool of disaggregated CPU servers for data preprocessing. }
	\label{fig:system_architecture_local_disagg_cpu}
	\vspace{-1.3em}
	\end{figure}

{\bf Hardware architecture.} While the software architecture of RecSys data
preprocessing provides high scalability,  a critical challenge remains
regarding \emph{how to allocate a sufficient number of CPU cores that provide
	high enough data preprocessing throughput that meets the throughput demands
		of GPU-side training?} Traditionally, model training and data preprocessing
	jobs are co-located within the same server node containing multiple GPUs (\fig{fig:system_architecture_local_disagg_cpu}(a))~\cite{
		vldb2021_mohan, tf_data}. As such, the
		performance of such co-located training pipeline is limited by how many CPU
		cores are available within the same server node (e.g., NVIDIA DGX system
			contains $8$ A100 GPUs and $128$ CPU cores, allowing a single GPU 
				to utilize $16$ CPU cores for data preprocessing), potentially suffering from 
		significant GPU underutilization when the aggregate CPU-side data
		preprocessing throughput underwhelms the  GPU-side
		training performance (detailed in \sect{sect:motivation}).

To address these challenges, Zhao et al.~\cite{dsi} and Audibert et
al.~\cite{disaggregation_google} proposed  server
disaggregation where a pool of CPU servers is reserved 
for data preprocessing (\fig{fig:system_architecture_local_disagg_cpu}(b)).
Disaggregating CPU servers allows a large pool of CPU cores to be elastically
allocated on-demand for preprocessing, effectively closing the performance
gap between preprocessing and model training~\cite{dsi,xdl}. However, such
design point incurs substantial deployment cost and power consumption due to
the large number of pooled CPU servers.

\section{Characterization and Motivation}
\label{sect:characterization_and_motivation}

\begin{table*}[t!]
  \centering
  \setlength{\tabcolsep}{3.0pt}
  \begin{tabular}{cc ccccc cccc}
  \toprule && \multicolumn{5}{c}{Data preprocessing configuration parameters} & \multicolumn{4}{c}{RecSys model architecture} \\
  \cmidrule(lr){3-7} \cmidrule(lr){8-11}
  \multicolumn{2}{c}{Type} & {\# Dense feats.} &  {\# Sparse feats.} & \thead{Avg. sparse \\feat. length} & \thead{\# Generated \\sparse feats.} & {Bucket size} & {Bottom MLP} & {Top MLP} & \thead{\# Tables} & \thead{Avg.\\ \# Embeddings}\\
  \midrule
  Public & RM1 & 13 & 26 & 1 (fixed) & 13 & 1024 & 512-256-128 & 1024-1024-512-256-1 & 39 & 500,000\\
  \midrule
  {\multirow{4}{*}{Synthetic}} & RM2 & 504 & 42 & 20 & 21 & 1024 & 512-256-128 & 1024-1024-512-256-1 & 63 & 500,000\\
  & RM3  & 504 & 42 & 20 & 42 & 1024 & 512-256-128 & 1024-1024-512-256-1 & 84 & 500,000\\
  & RM4  & 504 & 42 & 20 & 42 & 2048 & 512-256-128 & 1024-1024-512-256-1 & 84 & 500,000\\
  & RM5  & 504 & 42 & 20 & 42 & 4096 & 512-256-128 & 1024-1024-512-256-1 & 84 & 500,000\\
  \bottomrule
  \end{tabular}
  \caption{The RecSys training dataset configuration and the target model architecture. 
		RM1 is based on the public Criteo dataset~\cite{criteo:terabyte} and RM2-5 are synthetically generated models we created by referring to production-grade RecSys dataset's characteristics released by Meta~\cite{dsi}.}
  \label{tab:model}
  \vspace{-1.3em}
  \end{table*}

In this paper, we utilize the open-source RecSys data preprocessing library
TorchArrow~\cite{torcharrow:github} to conduct a workload characterization
study on state-of-the-art, CPU-centric data preprocessing systems.  We note
that there exists a significant disparity between the publicly available RecSys
dataset~\cite{criteo:terabyte} and the characteristics of a production-level
dataset mentioned by a recent work from Meta~\cite{dsi}. Specifically, compared
to the public dataset, production-level RecSys datasets contain a much larger
number of dense/sparse features and larger average sparse feature length.
Such discrepancy can undermine the primary objective of our study which is to characterize
			 state-of-the-art RecSys preprocessing.  Consequently, we scale up the
			 open-sourced Criteo dataset~\cite{criteo:terabyte} (referred to as RM1
					 in this paper) and develop four \emph{synthetic} RecSys models (RM2-5) based
			 on \cite{dsi} to better represent the properties of
			 production-level RecSys datasets. \tab{tab:model} summarizes the details
			 of our public/synthetic datasets and the RecSys models trained.
			 Our characterization is conducted with a training batch size of 8,192.
			 \sect{sect:methodology} further details our evaluation methodology.

\begin{figure}[t!]
\centering
\includegraphics[width=0.485\textwidth]{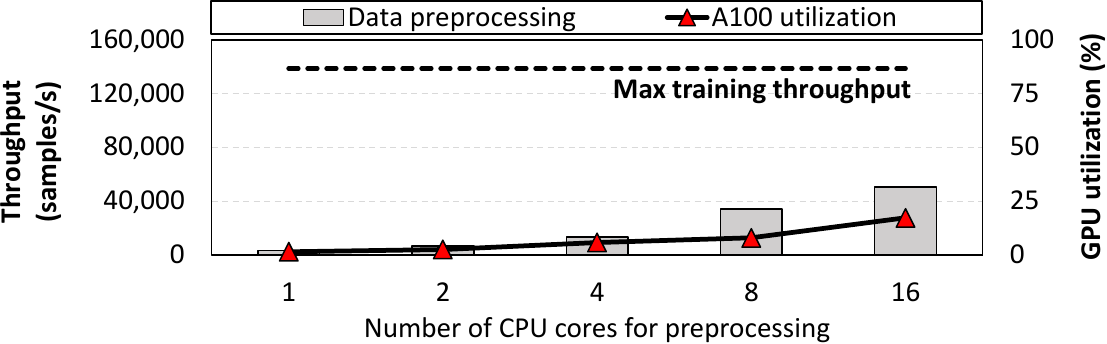} 
\caption{Effective preprocessing throughput (left axis) and the resulting GPU utilization (right axis) as a function of the number of CPU cores (i.e., number of preprocessing workers) utilized for preprocessing. The dotted line shows the upperbound, maximum training throughput achievable using a single NVIDIA A100 GPU (left axis), which assumes the GPU is seamlessly fed with sufficient amount of train-ready tensors without interruption. To measure GPU's utilization, we use the CUDA Profiling Tools Interface (CUPTI) library. The experiment is collected over the evaluation platform detailed in \sect{sect:methodology} using our synthetic model RM5.}
\label{fig:motivation_colocated_preproc}
\vspace{-1.3em}
\end{figure}

\subsection{Motivation}
\label{sect:motivation}

As discussed in \sect{sect:system_architecture_recsys}, the aggregate data
preprocessing throughput is strictly determined by how many CPU cores (i.e.,
		the number of data preprocessing workers) are utilized for
preprocessing.  Consider a co-location based RecSys training system
(\fig{fig:system_architecture_local_disagg_cpu}(a)) using a state-of-the-art DGX
server~\cite{a100_dgx} which contains $8$ A100 GPUs and $128$ CPU cores,
	allowing a single GPU to utilize $16$ CPU cores for data preprocessing.  In
	\fig{fig:motivation_colocated_preproc}, we scale up  the number of CPU cores
	for preprocessing (from $1$ to a maximum of $16$) and study its effect on
	data preprocessing throughput (left axis) and the percentage of execution
	time the A100 GPU is actually training the model (right axis).  We
	make the following two key observations from this experiment.  First, the
	preprocessing throughput increases almost linearly as a function of the number
	of CPU cores, achieving $15\times$ throughput improvement with $16$
	preprocessing workers vs. a single worker. Second, even with $16$
	preprocessing workers (i.e., the maximum number of CPU cores that can be
			allocated in co-located RecSys preprocessing), the GPU spends less than $20\%$ of its
	execution time actually conducting model training as the train-ready tensors
	are not being sufficiently supplied to the GPU. Consequently, the end-to-end
	training throughput gets bounded by the effective preprocessing
	throughput which is well below the maximum training throughput
	achievable (dotted line).

\begin{figure}[t]
\centering
\includegraphics[width=0.44\textwidth]{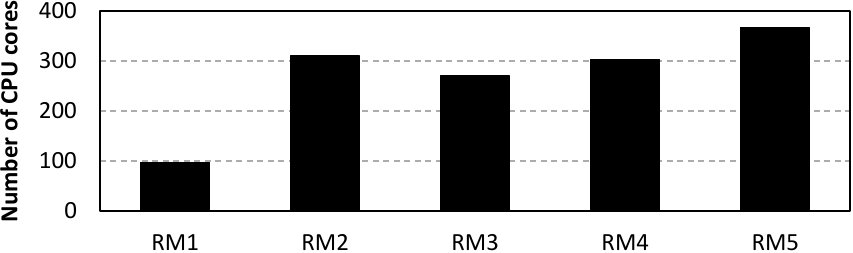} 
\caption{The number of CPU cores required for CPU-centric preprocessing to fully utilize a training node containing 8 A100 GPUs.} 
\label{fig:cpu_preproc_num_nodes}
\vspace{-1.3em}
\end{figure}

Server disaggregation for data preprocessing~\cite{dsi,disaggregation_google,xdl} is an effective
solution to close such wide performance gap, as it enables the allocation of
any number of CPU preprocessing workers as required by the GPU training stage
(\fig{fig:system_architecture_local_disagg_cpu}(b)).  In
\fig{fig:cpu_preproc_num_nodes}, we derive the number of CPU cores required in
the data preprocessing stage to sustain the  model
training stage's high throughput requirement. For production-level synthetic datasets with large number of features and sparse feature
lengths, several hundreds of CPU cores are required  ($367$ cores for RM5)
to sufficiently supply the train-ready tensors for an $8$ A100 GPU server node.
 It is important to note that hundreds to thousands of
	such production-level RecSys models are developed by 
	ML engineers, invoking numerous
concurrent			training jobs executed over
			several tens of thousands of high-performance GPUs across
			the	datacenter fleet~\cite{dsi,xdl}.
Such high demand for RecSys model training directly translates into substantial
cost and power consumption in maintaining the disaggregated CPU servers for data
preprocessing, e.g., Meta states that up to $60\%$
					of power consumption in RecSys training pipeline is dedicated to
					the storage and data ingestion pipeline of online CPU-centric data preprocessing~\cite{dsi}.

Given such, a key motivation of this work is to conduct a detailed characterization on
CPU-centric RecSys data preprocessing systems. In the remainder of this section, 
	we root-cause the system-level bottlenecks of existing solutions in search for
 a scalable and cost-effective preprocessing system for RecSys.

\subsection{Breakdown of End-to-End Data Preprocessing Time}

\label{sect:cpu_preproc_time}
\begin{figure}[t]
\centering
\includegraphics[width=0.47\textwidth]{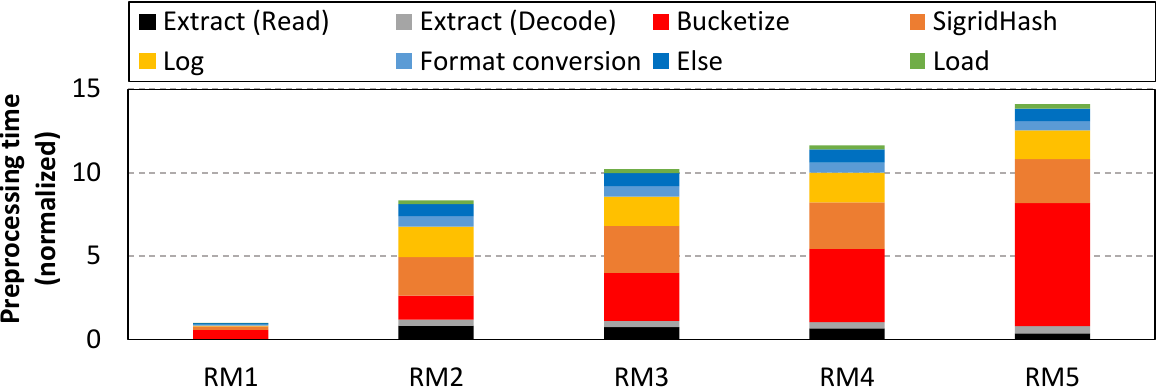} 
\caption{Latency to preprocess a single mini-batch input using a single preprocessing worker in the
	baseline CPU-centric system, broken into key steps of preprocessing. The ``Extract'' stage (\sect{sect:recsys_training_pipeline})  is further divided into (1) latency to fetch encoded raw feature
	data from the remote storage node (denoted as ``Extract (Read)'') and (2) latency spent decoding them (denoted as ``Extract (Decode)''). As depicted, data preprocessing is bounded by the compute-intensive feature generation and normalization operations, rather than I/O operations (``Extract (Read)''). All results are normalized to RM1.}
\label{fig:cpu_latency_breakdown}
\vspace{-1.3em}
\end{figure}

To identify the key bottlenecks in RecSys data preprocessing,
	 \fig{fig:cpu_latency_breakdown} first evaluates end-to-end latency to
	 preprocess a single training mini-batch during the ETL phase.  
	 Compared to the public RM1, the production-level RM2-5 
	 contain a larger number of dense and sparse features with larger average
	 sparse feature lengths. 
	 As such, these production-level 
	 models experience a substantial increase in
	 total preprocessing time where RM5 experiences the largest $14\times$ increase in latency.
	 Specifically, with the larger number of features in the RM2-5 models, both dense and sparse
	 feature normalization time (i.e., Log and SigridHash, respectively) account for up
	 to $55\%$ of its preprocessing time. 
	Additionally, these production-grade RM2-5 models
	 experience a notable increase in feature generation time (i.e., Bucketize)
	due to the large number of sparse features to generate and its large  bucket size.
	While RM3-5 all have the same number of sparse features to generate (at $42$), larger 
	bucket sizes (from $1024$ to $4096$ in RM3 to RM5, $m$ in \algo{algo:bucketize}) incur longer feature generation time because the
	latency to search the bucket ID of each feature value increases. Conversely, despite RM1-3 all having the same bucket size at $1024$,
	the larger number of sparse features to generate (from $13$ to $42$ in RM1 to RM3) leads to larger feature generation time as
	well.

	Overall, feature generation (Bucketize) and feature normalization (SigridHash and Log) collectively account
	for an average $79\%$ of the data preprocessing time and become the most
	significant performance limiter in RecSys preprocessing.

\subsection{Analysis on Performance-Limiting Operations}
\label{sect:characterization}

\begin{figure}[t]
\centering
\includegraphics[width=0.47\textwidth]{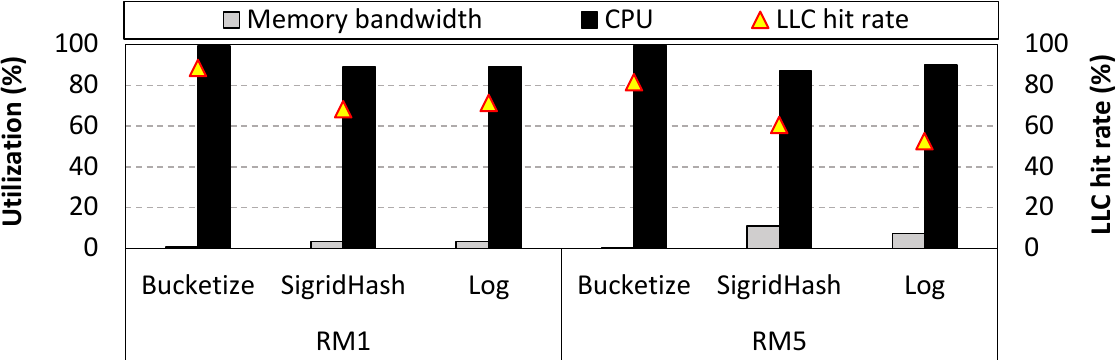} 
\caption{CPU and memory bandwidth utilization (left) and LLC hit rate (right) during the execution of feature generation/normalization operations for the public model (RM1) and the production-scale model (RM5). We utilize Linux \texttt{perf} for our evaluation.}
\label{fig:characterization_preproc_ops}
\vspace{-1.3em}
\end{figure}

\fig{fig:characterization_preproc_ops} shows the CPU and memory bandwidth
	utilization and the last-level cache (LLC) hit rate during the execution of
	the three key performance-limiting operations (i.e., Bucketize, SigridHash,
			Log) in RM1 and RM5, respectively.  Our
	analysis reveals the following key insights.  First thing to note is that all
	three operations exhibit high CPU utilization with relatively low memory
	bandwidth utilization.  RM5 in particular achieves increased memory
		bandwidth utilization as it has more features to generate and normalize
			compared to RM1. Nonetheless, the memory bandwidth utility of RM5 still
			remains under $15\%$ of the maximum $281.6$ GB/sec of memory throughput,
			exhibiting a compute-bound behavior.  While the feature generation and
			normalization operations require large amount of data accesses, the
				actual working set it touches upon is relatively small. In RM1 and RM5,
				it amounts to as small as several tens of KBs to at most tens of
					MBs. For instance, the
					Bucketize operation involves sharding features based on the
					predefined bucket range whose active working set fits well within
					on-chip caches, leading to an LLC hit rate of $85\%$.

\subsection{Our Goal: Scalable $\&$ Cost-Effective Preprocessing}

	Overall, we conclude that current CPU-centric RecSys preprocessing systems
	are bounded by the level of computation power available with CPUs,
	failing to fully reap out the inter-/intra-feature parallelism inherent in
	preprocessing. Although disaggregating a pool of CPUs for
	data preprocessing can help meet the computation demands of
	preprocessing, it requires high deployment cost and high power consumption.
	We argue that the abundant inter-/intra-feature parallelism in data
	preprocessing is well-suited for domain-specific acceleration, proposing an
	\emph{accelerated} computing system for RecSys data preprocessing which is
	scalable and cost-effective. 

\section{\ours: An In-Storage ``Pre''processing Architecture for RecSys Training}
\label{sect:proposed}

We present \ours ({\bf Pre}processing in-{\bf Sto}rage), our In-Storage Processing (ISP) based data preprocessing system for RecSys training. 
Our proposed system offloads the
	time-consuming feature generation and normalization operations to an
	accelerator that is tightly coupled with the storage system.

\begin{figure}[t!]
	\centering
	\includegraphics[width=0.48\textwidth]{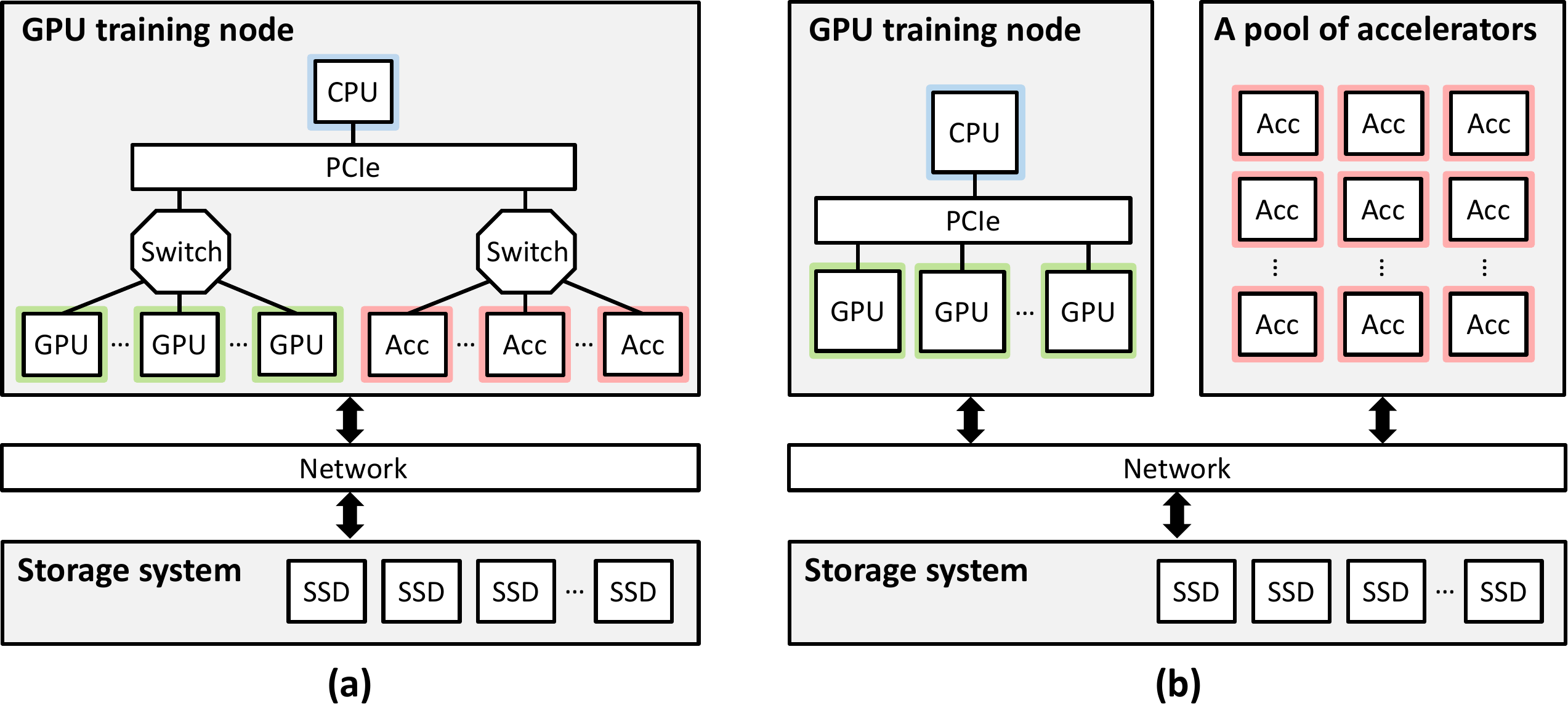} 
	\caption{(a) Preprocessing accelerators co-located with GPUs and (b) disaggregated pool of preprocessing accelerators.}
	\label{fig:system_designs}
  \vspace{-1.3em}
  \end{figure}

\subsection{System Design Considerations}
\label{sect:dse}

Our key proposition is to employ accelerated computing for
data preprocessing, so an important question to be answered is \emph{where} our
accelerator(s) should be deployed within the system architecture. Below we elaborate
on the two possible design points that retrofit our data preprocessing accelerator within 
conventional co-located and disaggregated server architectures.

{\bf Preprocessing accelerators co-located with GPUs.}
\fig{fig:system_designs}(a) illustrates our first design point, a scale-``up''
server architecture employing a PCIe switch to co-locate the preprocessing
accelerators with the GPUs.  When the number of co-located accelerators is
large enough to meet the GPU's training demands, this design point can obviate
the need for disaggregated CPU servers dedicated to preprocessing. However, a
critical limitation of such scale-up server is its \emph{scalability} because
the aggregate preprocessing throughput is still bounded by the total number of
accelerators co-located within this scale-up server. As such, for large-scale
RecSys models whose data preprocessing demands exceed the preprocessing
throughput available with co-located accelerators, the GPU training workers will
still suffer from idle periods. Another key concern with this approach is that
both the accelerator-side data preprocessing workers and the GPU-side training
workers all time-share the PCIe bus to communicate with the host CPU.
Because preprocessing workers and training workers  concurrently
execute in a training pipeline, the PCIe bus can become a performance
hotspot under such unbalanced system architecture. Given such critical
limitations, we conclude that such scale-up solution is impractical.

{\bf Disaggregated pool of preprocessing accelerators.} To address the
scalability issue in our scale-up server design, an alternative solution would
be to utilize our preprocessing accelerator as a drop-in replacement of CPUs in
the disaggregated preprocessing CPU pool. As shown in
\fig{fig:system_designs}(b), this design point effectively provides a
disaggregated pool of preprocessing accelerators to the GPU training workers.  By
decoupling training jobs from preprocessing jobs over distinct server pools, the
optimal number of preprocessing accelerators to allocate that meets a target
training job's throughput demands can be determined dynamically, providing
high scalability. Furthermore, this design point can better exploit
inter-/intra-feature parallelism using domain-specific acceleration for high
efficiency.  However, similar to the baseline CPU-centric disaggregated
preprocessing, server disaggregation still incurs substantial deployment cost
and high TCO.

\subsection{Proposed Approach: In-Storage Data Preprocessing}
\label{sect:proposed_approach_isp}

\begin{figure}[t!]
	\centering
	\includegraphics[width=0.40\textwidth]{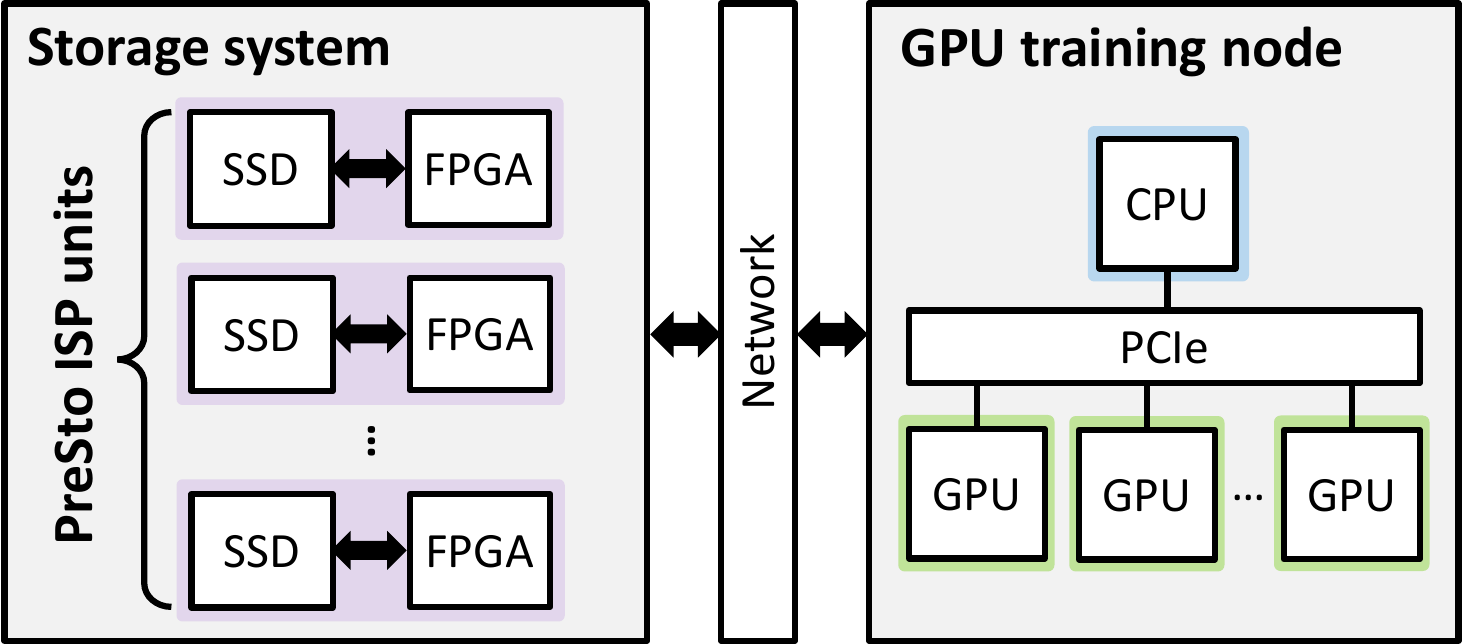} 
	\caption{System architecture of PreSto.}
	\label{fig:proposed_system}
  \vspace{-1.3em}
  \end{figure}

{\bf Hardware architecture.} Due to the aforementioned challenges of co-located
and disaggregated accelerator systems, our proposed system employs ISP (in-storage
		processing) architectures for data preprocessing, i.e., \emph{in-storage
	``pre''processing}.  As shown in
	\fig{fig:proposed_system}, this approach utilizes ISP devices like
	SmartSSDs~\cite{smartssd} to directly replace conventional SSD cards. A
	SmartSSD tightly couples a normal SSD with a lightweight FPGA device within
	the NVMe U.2 form factor. This allows SmartSSDs to become a drop-in
	replacement for normal SSDs while still staying within the NVMe's $25$ Watts power envelope\footnote{A
		high-end FPGA card like Xilinx U280~\cite{u280} which has a TDP of 225
			Watts cannot be utilized for a U.2 SmartSSD card.  }. As such, a SmartSSD
			can utilize its local FPGA device to implement our preprocessing accelerator
			right next to the local SSD where the raw feature data is stored.  Such design
			point effectively tackles the system challenges of both co-located and
			disaggregated preprocessing accelerators as follows. 

  \begin{figure}[t]
    \centering
    \includegraphics[width=0.40\textwidth]{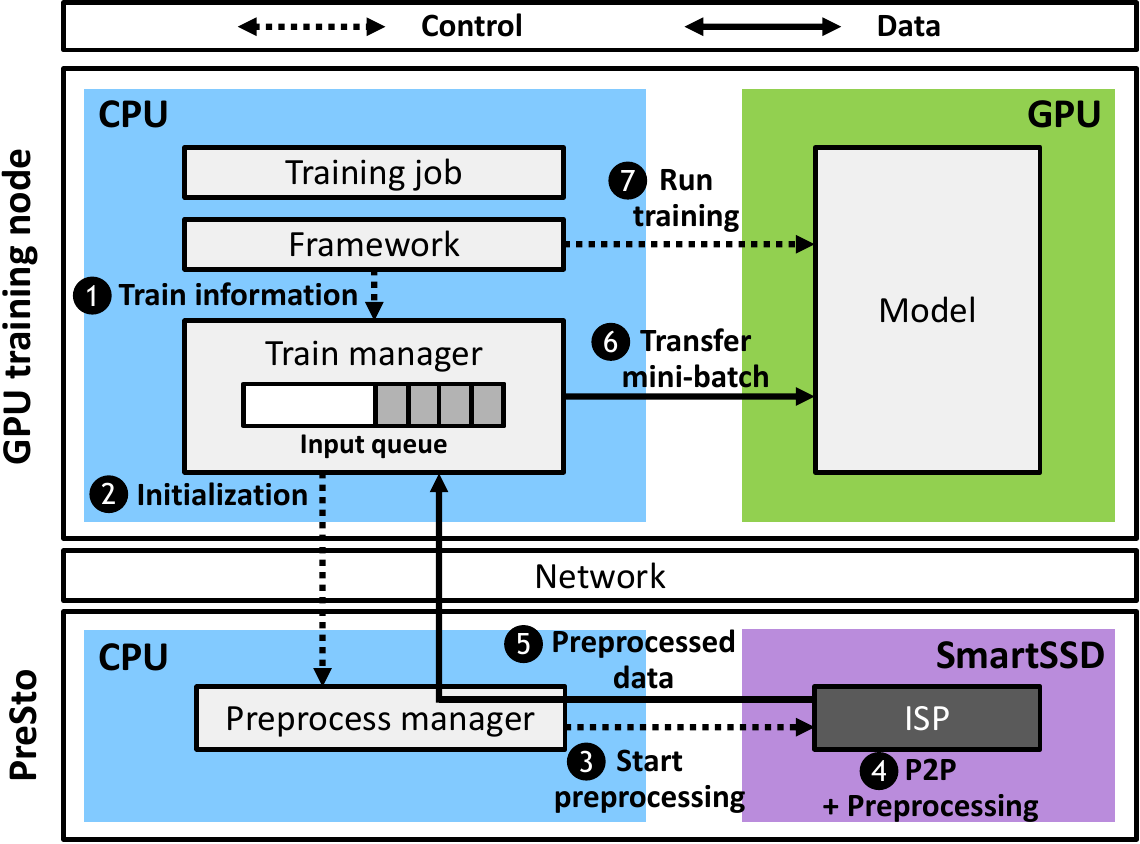} 
    \caption{Major components of \ours software system and key steps undertaken during end-to-end RecSys training.}
    \label{fig:proposed_software_system}
  \vspace{-1.3em}
  \end{figure}

\begin{enumerate}
\item {\bf Scalability.} As discussed in \fig{fig:recsys_pipeline}, the tabular
raw feature data subject for preprocessing is stored as columnar files. A group
of rows within the tabular data is sharded into partitions and different
partitions are stored as independent columnar files in a distributed storage
system (e.g., $2$ columnar files stored in two SSDs in
		\fig{fig:recsys_pipeline}). 
		While multiple blocks that constitute a single columnar file can further be distributed across
multiple storage devices, state-of-the-art file systems for RecSys (e.g., Meta's
		Tectonic file system\cite{tectonic}) store all the blocks in a given partition
contiguously within a single storage device. This ensures that all preprocessing operations can be conducted locally within a SmartSSD because all transformation
operations conducted within a mini-batch (i.e., partition) are executed locally without any dependencies to other mini-batches (i.e., partitions in other columnar files).
Consequently, in our proposed system, the overall
preprocessing throughput can scale proportionally with the number of SmartSSDs
allocated to a preprocessing worker targeting a given training job.  
A training job
							 with a target preprocessing throughput in mind is first
							 allocated with the appropriate number of SmartSSDs (detailed
									 later in this subsection), one that is large enough to
							 sustain the training stage's preprocessing need. We then have
each SmartSSD independently extract raw feature data from its local SSD, which
is immediately P2P transferred over to the local FPGA device for on-the-fly
preprocessing. Because multiple SmartSSDs (i.e., multiple preprocessing
		workers) concurrently conduct preprocessing and generate mini-batch inputs,
	our \ours ISP units provide highly scalable preprocessing service.

		\item {\bf Efficiency.} In our proposed system, all preprocessing
		operations are conducted \emph{locally} within the storage system. This is
		because the SmartSSD's FPGA accelerator can extract the raw feature data
		directly from its local SSD using P2P data transfers, obviating the need
		for a disaggregated accelerator server pool with high maintenance cost.
		Such design decision also helps eliminate the communication overhead
		associated with disaggregated accelerator server designs
		(\fig{fig:system_designs}(b)), which requires the raw feature data to be
		transferred from the storage system to the remote accelerator pool.  It is
		also worth pointing out that SmartSSDs can seamlessly be deployed within
		the power constraints of the baseline distributed storage system, all
		thanks to the use of commodity devices that operate within the NVMe SSD's
		power budget (less than 25 watts per device~\cite{csd_power_cidr_2021}).
		Consequently, \ours is minimally intrusive to existing hardware
		infrastructure while maintaining power-efficiency via accelerated
		computing.

\end{enumerate}

{\bf Software architecture.} \fig{fig:proposed_software_system} provides a high-level
overview of our software architecture. The two key components of our software
system are the train manager and the preprocess manager. The train manager
is implemented as part of the training worker process whose main role is to manage the
end-to-end model training job, from data preprocessing to model training.
That is, it requests the mini-batch inputs (i.e., train-ready tensors) from the
storage system and once the mini-batch inputs are returned, they are forwarded to the GPU for
model training.  The preprocess manager is in charge of spawning and managing the
actual preprocessing workers using the SmartSSD devices. Once the
mini-batch inputs are ready, the preprocess manager returns them back
to the train manager. Below we summarize the major steps undertaken during
the end-to-end RecSys training process.

\begin{enumerate}

\item When a training job is launched by TorchRec~\cite{torchrec},  the train
manager receives important information about the target training job (e.g.,
				model configuration for training/preprocessing, mini-batch size, and other
				metadata) and goes through several boot-strapping procedures in preparation for
model training. These include the 
allocation of an input queue to store the mini-batch inputs designated for model
				training and a Remote Procedure Call (RPC) initialization (step
						\ding{182} in \fig{fig:proposed_software_system}).
    
\item Using the training job information, the train manager measures the
maximum training throughput achievable with GPUs for the given training job.
This is done by supplying the GPUs with dummy mini-batch inputs and
stress-testing their highest sustainable throughput.  This process only takes
tens of seconds, the overhead of which is amortized over the several hours/days worth of training
time. The train manager then initializes the preprocess manager, sending
information relevant to the preprocessing jobs (e.g., configuration
		parameters of preprocessing). One of the important information that is
forwarded to the preprocess manager is GPU's maximum training throughput ($T$),
					as it determines the level of preprocessing throughput that must be
					sustained in order to fully utilize the GPUs for model training. As
					such, the preprocess manager measures offline the maximum preprocessing
					throughput delivered with a single SmartSSD device under the given
					preprocessing configuration ($P$). By dividing the maximum training
					throughput with the per-SmartSSD preprocessing throughput ($T$/$P$),
					the preprocess manager derives the number of SmartSSD devices
					that need to be allocated to fully saturate the GPUs (step \ding{183}).
    
\item The preprocess manager launches the necessary number of preprocessing
workers based on the derived number of SmartSSD ($T$/$P$) (step \ding{184}). As each
preprocessing worker independently generates mini-batch inputs locally within the
SmartSSD, each device fetches its share of raw feature data from the local SSD
and transfers them P2P to the FPGA to conduct preprocessing on-the-fly
 (step \ding{185}).

\item Once the mini-batch inputs are ready, they are converted into a train-ready tensor
format, as required by TorchRec, and copied over to the input queue within the
train manager (step \ding{186}).
    
\item Finally, the train manager transfers the mini-batch from its input queue to
the GPU and kicks off the model training process (step \ding{187} and \ding{188}). 
Because multiple SmartSSDs are concurrently preprocessing raw features and replenishing
the train manager's input queue, it ensures that the GPU is constantly supplied with
sufficient amount of mini-batch inputs to consume.

\end{enumerate}

\begin{figure}
  \centering
  \includegraphics[width=0.41\textwidth]{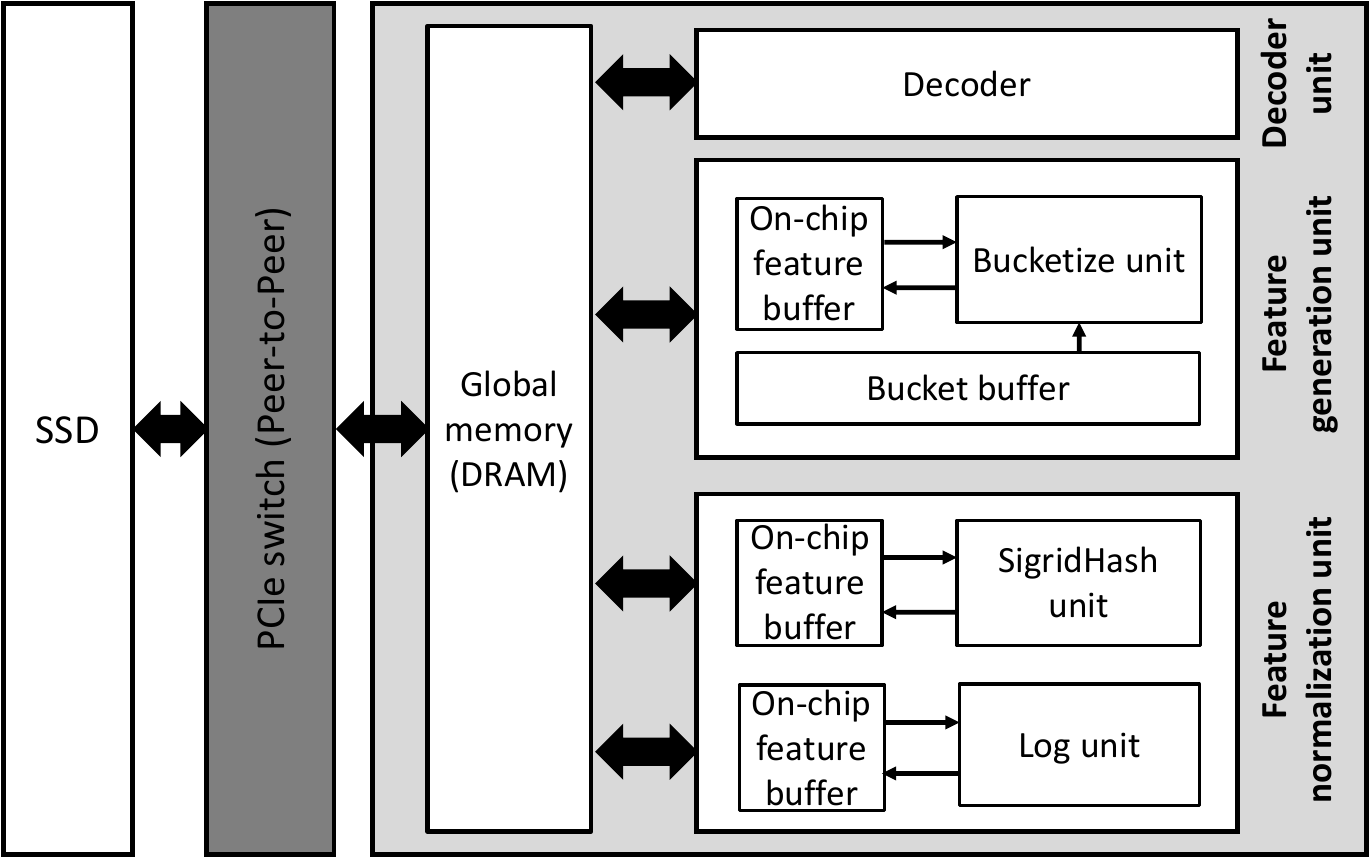} 
  \caption{\ours accelerator microarchitecture.}
  \label{fig:isp_architecture}
  \vspace{-1.3em}
\end{figure}

\subsection{Data Preprocessing Accelerator Microarchitecture}
\label{sect:microarch_isp_unit}

The design objective of our \ours accelerator is to maximally
exploit inter/intra-feature parallelism inherent in RecSys data preprocessing.
As depicted in \fig{fig:isp_architecture}, we use the custom logic
	 within the FPGA to implement a hardwired decoder unit,
	 feature generation units, and feature normalization units.  Each unit is equipped with the essential hardware logic
	 tailored to the following operations: ``Decoder'' for columnar file decoding
	 (our columnar files assume the Apache Parquet file format~\cite{apache_parquet}), ``Bucketize'' for feature generation, and
	 ``SigridHash'' as well as ``Log'' for feature normalization.
		 To maximally
	 exploit inter-/intra-feature parallelism, we employ the following
	 design optimizations.  First, to exploit inter-feature parallelism, we deploy
	 multiple processing elements dedicated to each individual feature, directly
	 connected to the off-chip-memory interface to fully utilize the bandwidth of
	 global memory (DRAM).  Second, to exploit intra-feature parallelism, each
	 processing element employs double-buffering to overlap the next feature value's data
	 fetch operation with the current feature value's generation and normalization operations. That
	 is, once a portion of an input feature is fetched on-chip, its
	 transformation is immediately executed while the next feature value is
	 concurrently being fetched from the off-chip memory.  

\section{Methodology} 
\label{sect:methodology}

\subsection{Benchmarks} 

The majority of current academic research on RecSys utilizes the Criteo
dataset~\cite{criteo:terabyte}, which is the largest publicly available dataset
for RecSys. The Criteo dataset consists of 13 dense and 26 sparse features,
		with a fixed feature length of 1 for each sparse feature. According to
		recent work from Meta~\cite{dsi}, production-level RecSys models can amount
		to 504 dense and 42 sparse features with an average sparse feature length of 20,
		much larger than the public Criteo dataset. To narrow this gap in our evaluation,
	we additionally construct four
		synthetic RecSys models in accordance with \cite{dsi}, expanding the existing
		features of the Criteo dataset to better cover the evaluation space of
		production-level RecSys models with large number of dense/sparse features.
		\tab{tab:model} details the configuration of the five RecSys models and their training
		datasets we evaluate.

\subsection{Experimental Setup}
\label{sect:method_simulator}

{\bf Hardware.} Exploring \ours in a production-level training
pipeline that accurately reflects industry's large-scale disaggregated CPU
servers, multi-node/multi-GPU systems, and a distributed storage array
integrated with SmartSSDs is challenging at the academic research level for
several reasons. Aside from many undisclosed details of hyperscaler's
production ML infrastructure, the unavailability of SmartSSDs in cloud services
like Amazon AWS~\cite{aws} rendered our experiments to employ the following
methodology. We first demonstrate \ours's advantages  in real systems by
constructing a small-scale, proof-of-concept (PoC) prototype of \ours using
commodity CPU/GPU/SmartSSD devices.  We then develop an analytical 
model that estimates \ours's performance at large-scale by utilizing real
measurements from our PoC prototype.

	\begin{table}[t]
  \centering
  \setlength{\tabcolsep}{3.5pt}
  \begin{tabular}{cccccc}
  \toprule
  Unit & LUT & REG & BRAM & URAM & DSP \\
  \midrule
  Decode & 18.84\% & 8.49\% & 25.08\% & 0.00\% & 0.00\% \\
  \midrule
  Bucketize & 7.88\% & 4.28\% & 6.19\% & 27.59\% & 0.00\% \\
  \midrule
  SigridHash & 23.11\% & 12.47\% & 11.89\% & 0.00\% & 19.19\% \\
  \midrule
  Log & 4.18\% & 2.79\% & 4.89\% & 0.00\% & 10.62\% \\
  \midrule
  Total & 54.02\% & 28.03\% & 48.05\% & 27.59\% & 29.81\% \\
  \bottomrule
  \end{tabular}
  \caption{FPGA resource utilization of \ours's preprocessing accelerator. Decode, Bucketize, SigridHash, and Log units are synthesized with an operating frequency of 223MHz.}
  \label{tab:resrc_util}
  \vspace{-1.3em}
  \end{table}

					 At a high-level, our PoC prototype includes three major components:
					 (1) a storage node (with and without a SmartSSD to model \ours (with
								 SmartSSD) and baseline storage system (without SmartSSD)), (2)
					 a GPU training node, and (3) a pool of multiple CPU nodes for
					 preprocessing (to model baseline disaggregated CPU preprocessing
							 service), which communicate over a network using $10$ Gbps
					 Ethernet.  Both the storage node and the pool of CPU nodes for
					 preprocessing are designed using a total of three two-socket Intel
					 Xeon Gold 6242 CPU nodes (32 CPU cores per node) where one node is
					 used as the storage node and the other two nodes are utilized as a
					 remote pool for data preprocessing (maximum 2$\times$32$=$64 CPU
							 cores available for data preprocessing).  The GPU training node
					 contains AMD EPYC 7502 CPU connected with a single NVIDIA A100 GPU.
					 When evaluating \ours, we add a single SmartSSD card~\cite{smartssd} to the
					 storage node which handles data preprocessing locally within the
					 storage node.  \ours's preprocessing accelerator is designed using
					 Xilinx Vitis HLS 2022.2 whose resource utilization is summarized in
					 \tab{tab:resrc_util}. 

					 As for the analytical model we developed for
					 large-scale performance estimations, we utilize the observations
					 made from our characterization study in
					 \sect{sect:characterization_and_motivation} where data preprocessing
					 operations are embarrassingly parallel and exhibit high
					 scalability. Because end-to-end training performance as well as its
	data preprocessing performance is throughput-bound rather than latency-bound,
	our analytical performance model assumes that the preprocessing throughput
	measured from our real PoC prototype (i.e., the preprocessing throughput measured
			with a single CPU core (baseline) and a single SmartSSD (\ours)) scales proportionally with the number of
	CPU cores allocated (baseline) or the number of SmartSSDs allocated (\ours)
	for data preprocessing.

{\bf Software.} Our end-to-end RecSys training pipeline is implemented using
TorchArrow (v0.1.0)~\cite{torcharrow:github} for data preprocessing and TorchRec
(v0.3.2)~\cite{torchrec} for model training, assuming a mini-batch size
of 8,192.  We assume the Apache Parquet file format~\cite{apache_parquet} when the columnar
raw feature data is stored in our storage system.  Both baseline
CPU-centric and \ours preprocessing system communicate with the GPU training
node using 
the PyTorch RPC
API~\cite{pytorch_rpc}.  We use
	Xilinx Runtime library to manage \ours's FPGA device. 

\subsection{Evaluation Methods}
\label{sect:method_metric}

{\bf Power measurement.} When measuring the power consumption of the CPU-based
storage node as well as the disaggregated preprocessing nodes, we measure its
system-level power consumption using Intel Performance Counter Monitor
(PCM)~\cite{intel_pcm}.  The Xilinx Vivado~\cite{xilinx_vivado} and NVIDIA
System Management Interface (nvidia-smi) are used when measuring the power
of \ours's FPGA accelerator and the GPU, respectively.

{\bf Cost-efficiency.} 
To quantify the cost-effectiveness of \ours, we also evaluate cost-efficiency using the evaluation metric
suggested in \cite{liu_e3,dscs_arxiv} as summarized below:

\[\text{Cost-efficiency} = \frac{Throughput\times Duration}{\text{CapEx}+\text{OpEx}}\]
\[\text{where OpEx} = \sum(Power\times Duration\times Electricity)\]

CapEx (\$) refers to the one-time capital expenditure required to purchase and
establish the hardware platform components. OpEx (\$) represents the operating
expenditure of this hardware platform.  To determine CapEx, we utilize the cost
information obtained from the respective company
website~\cite{dell_r640,smartssd}.  OpEx is derived using the power consumed by
the hardware components ($Power$), the active duration of each hardware
component ($Duration$, a period of 3 years~\cite{dscs_arxiv,barroso2019datacenter}), and the average price of electricity
($Electricity$, \$0.0733/kWh~\cite{liu_e3,dscs_arxiv}). It is worth pointing out
that the numerator value to calculate cost-efficiency
($Throughput$$\times$$Duration$) is identical for both baseline disaggregated
CPU preprocessing and \ours: both baseline and our proposal
can sustain the throughput demands of GPU's training stage, so $Throughput$
and $Duration$ are constant values. Therefore, the difference in
cost-efficiency is determined by (CapEx+OpEx).
\section{Evaluation}
\label{sect:evaluation}

We first demonstrate \ours's merits using our PoC prototype
(\sect{sect:eval_poc}). We then utilize our analytical model
(\sect{sect:methodology}) to estimate \ours's effect on performance,
	energy-efficiency, and TCO at larger scale (\sect{sect:eval_large_scale}).  In
	the rest of this section, the baseline CPU-centric preprocessing system
	assumes the disaggregated CPU server design (denoted ``\baseline''), one which
	is equipped with a maximum of $64$ CPU cores in our small-scale PoC
	prototype.

\subsection{Performance and Cost-Effectiveness of \ours (PoC)}
\label{sect:eval_poc}

\begin{figure}[t]
  \centering
  \includegraphics[width=0.49\textwidth]{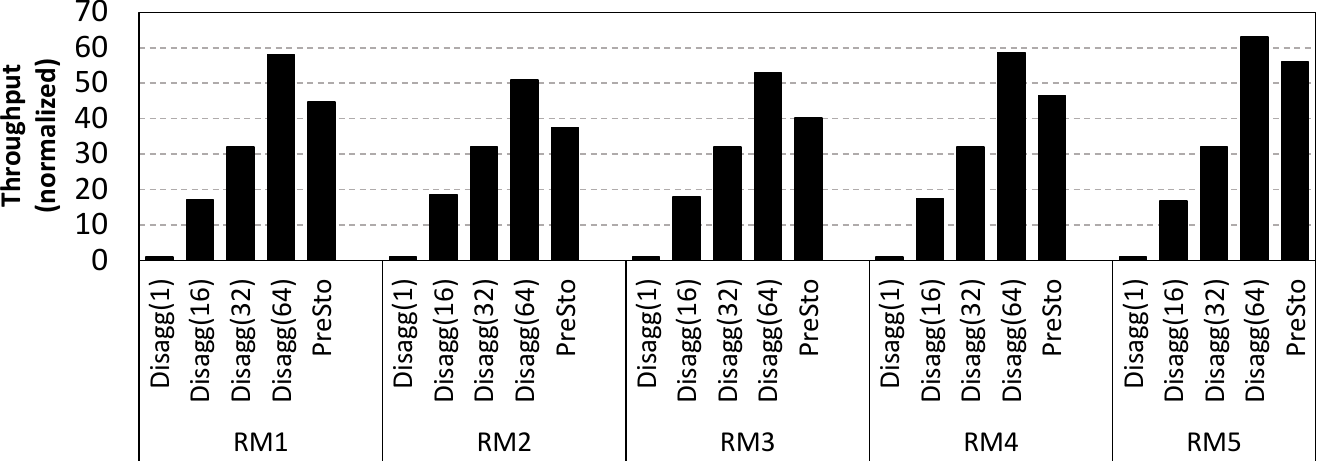} 
  \caption{Preprocessing throughput of \ours (single SmartSSD) vs. \baseline. Disagg($N$) is a design point executing with $N$ preprocessing workers using $N$ CPU cores. Results are normalized to Disagg(1).}
  \vspace{-1.0em}
  \label{fig:smartssd_throughput}
\end{figure}

{\bf Throughput.} \fig{fig:smartssd_throughput} compares the data
preprocessing throughput of \ours vs. \baseline. As depicted, 
a single SmartSSD device (\ours) consistently outperforms \baseline even with $32$ CPU
cores (i.e., a single CPU node) and demonstrates the benefits of our ISP solution.
Since \baseline's performance scales well to the number of CPU cores (workers)
utilized, allocating more CPU cores can still match the throughput provided with \ours albeit at a proportional increase in cost (e.g, Disagg(64) using two CPU nodes ($64$ cores) is able to slightly outperform \ours by average $27\%$ but with $2\times$ higher cost). 

\begin{figure}[t]
  \centering
  \includegraphics[width=0.49\textwidth]{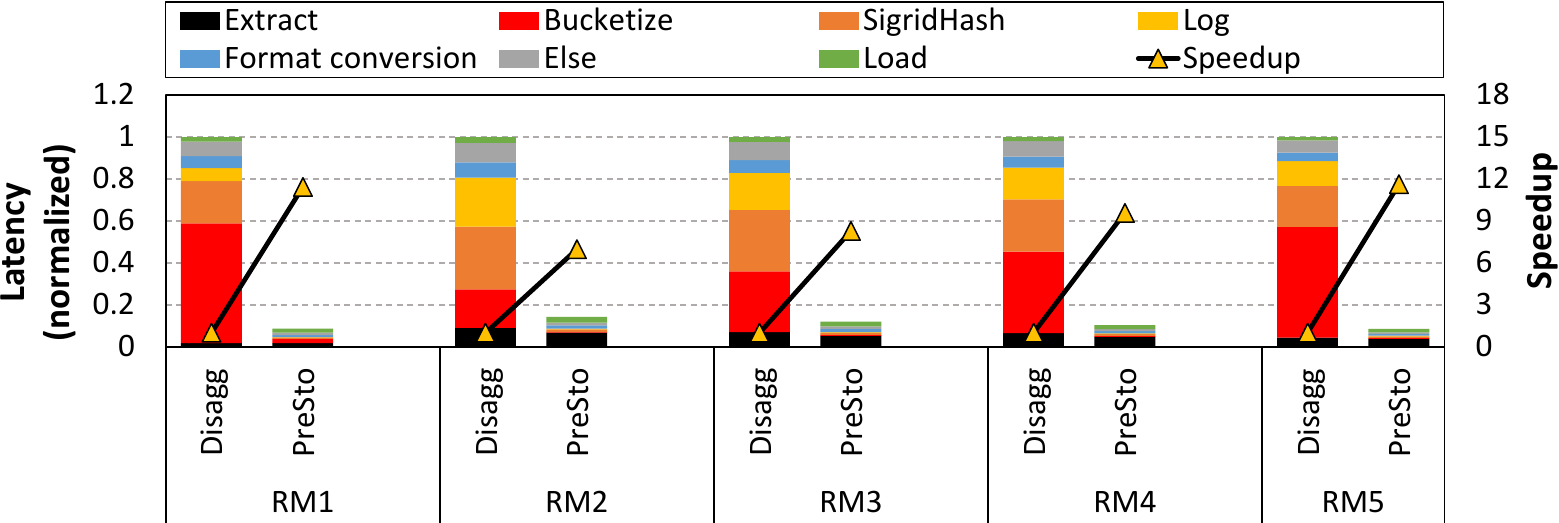} 
  \caption{(Left axis) Data preprocessing time to generate
		a single mini-batch using a single preprocessing worker. Latency is broken down into key steps of data preprocessing.  
  (Right axis) \ours's end-to-end speedup for preprocessing. All results are normalized to \baseline.}
  \vspace{-0.8em}
  \label{fig:smartssd_latency_breakdown}
\end{figure}

\begin{figure}[t]
  \centering
  \includegraphics[width=0.43\textwidth]{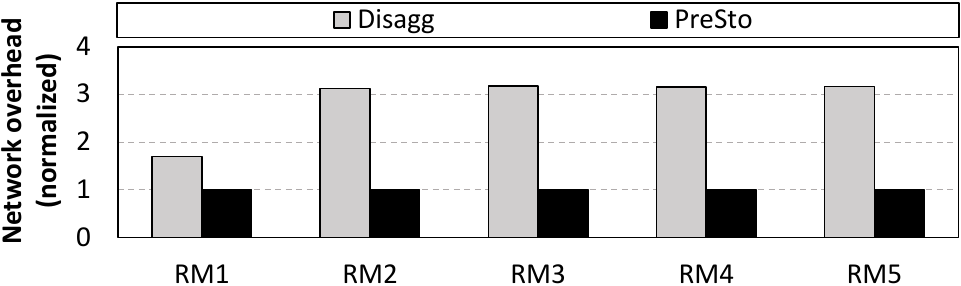} 
  \caption{Aggregate latency incurred during any RPC calls executed for inter-node communication during the course of data preprocessing. 
}
  \vspace{-0.8em}
  \label{fig:network_traffic}
\end{figure}

{\bf Latency.} To better highlight where \ours's speedup comes
from, \fig{fig:smartssd_latency_breakdown} compares the latency  to
generate a single mini-batch input using a single preprocessing worker using
\baseline and \ours.  The latency breakdown focuses on the key
steps undertaken during preprocessing.
In the baseline \baseline, the ``Extract'' step includes time to fetch encoded raw
feature data from the remote storage node and decode them.  With \ours,
				the ``Extract'' step includes the P2P data transfer of the encoded raw
				feature data from local SSD to the FPGA which is immediately followed by their decoding using our
				dedicated decoder unit. Because the decoding algorithm is less
				parallelizable than feature generation and normalization operations,
				the reduction in the ``Extract'' step's execution time is less pronounced, rendering
				this step to account for an average 40.8\% of the total
				preprocessing time of \ours. Nonetheless, \ours provides significant
				improvements in the performance of feature generation (Bucketize) and
				normalization (SigridHash, Log), achieving an average $9.6\times$
					(maximum $11.6\times$) reduction in end-to-end preprocessing time.
					These results highlight the benefits of \ours's domain-specific acceleration
					using RecSys preprocessing's inter-/intra-feature parallelism.

{\bf Data movements.} Another key benefit provided with \ours is that all data
preprocessing operations are conducted locally within the storage node, unlike
\baseline which needs to explicitly copy data in (the raw data to be
		preprocessed) and out (the train-ready tensors) of the disaggregated CPU
nodes for data preprocessing. Because our small-scale PoC prototype evaluates a
single training job in a highly controlled, isolated setting, \baseline's RPC
communication time to read out the raw feature data from the remote storage
node and copy into the disaggregated CPU nodes accounts for a relatively small
portion of the end-to-end preprocessing time (but still accounting for
		$9.1\%$ in RM2 under \baseline in
		\fig{fig:smartssd_latency_breakdown}).  Since real-world datacenter fleets
concurrently handle a large number of training jobs, all of which time-share
the datacenter network, \ours's ISP capability can be beneficial in alleviating
the preprocessing operation's pressure on network communications.  In
\fig{fig:network_traffic}, we show the aggregate latency incurred during any
RPC calls executed for inter-node data movements during the course of data preprocessing.
Unlike \baseline which incurs additional latency
when the preprocessing worker copies raw feature data from the remote storage
to the disaggregated CPU nodes, our \ours can completely
eliminate such performance overhead. This leads \ours to provide a $2.9\times$ reduction in
RPC-invoked inter-node communication time.

\begin{figure}[t]
	\centering
	\includegraphics[width=0.41\textwidth]{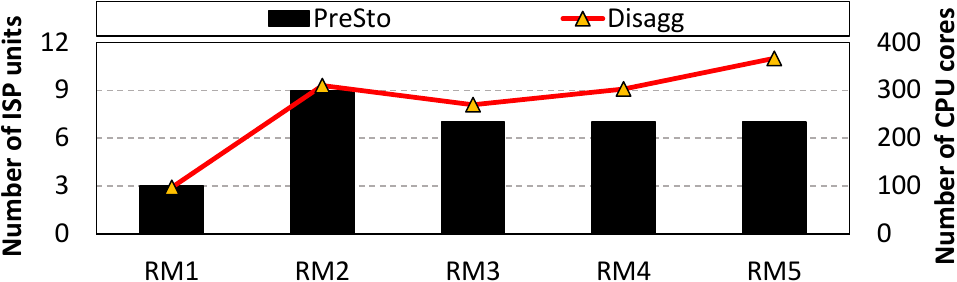} 
	\caption{The number of ISP units (left axis) and CPU cores (right axis) required for \ours and \baseline to sustain a single multi-GPU server node containing $8$ A100 GPUs.}
	\vspace{-0.8em}
	\label{fig:smartssd_preproc_num_devices}
  \end{figure}

\subsection{\ours's Effect on Energy-Efficiency and TCO}
\label{sect:eval_large_scale}

We now evaluate \ours's effect on performance/Watt (energy-efficiency) and performance/\$ (TCO) by utilizing our analytical model for large-scale experiments (\sect{sect:method_simulator}).
  
  \begin{figure}[t]
	\centering
	\includegraphics[width=0.49\textwidth]{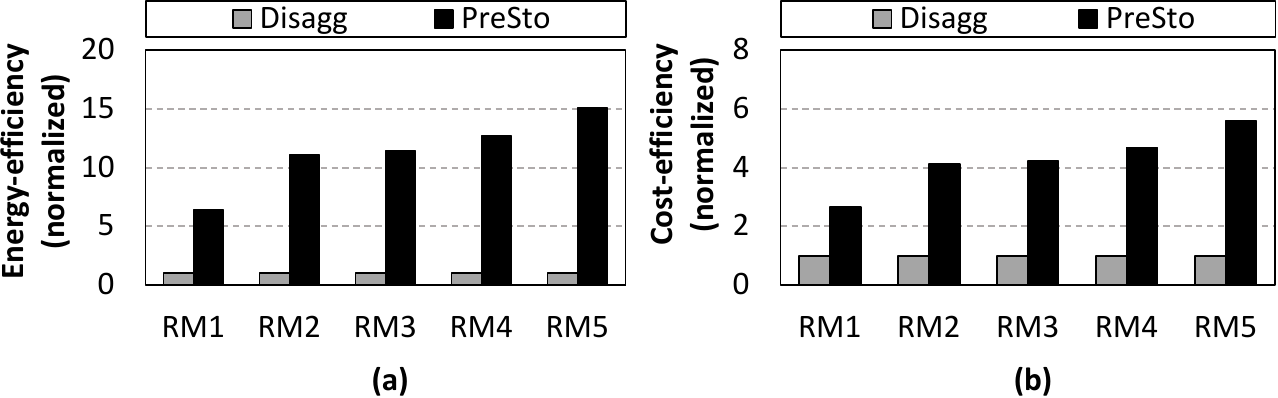} 
	\caption{(a) Energy-efficiency and (b) cost-efficiency. Power consumption is measured using Intel's Performance Counter Monitor (PCM) and Xilinx Vivado. \sect{sect:method_metric} details our methodology.}
	\vspace{-1.3em}
	\label{fig:energy_cost_efficiency}
  \end{figure}

{\bf Energy-efficiency.} As discussed in \fig{fig:cpu_preproc_num_nodes},
	baseline \baseline requires significant number of CPU cores for data
	preprocessing (e.g., $367$ cores for RM5), which translates into significant
	power consumption and high deployment cost. To quantitatively demonstrate
	\ours's effectiveness in energy and cost reduction, we evaluate how many ISP
	units (i.e., the number of SmartSSD cards) are required to match the
	preprocessing demands of a multi-GPU server containing $8$ GPUs
	(\fig{fig:smartssd_preproc_num_devices}).  Remarkably, to match such  high
	GPU training throughput demand, \ours only requires a maximum of $9$
	ISP units which incur (9$\times$25)$=$225 Watts of worst-case power
	consumption ($25$ Watts TDP per each SmartSSD card). \baseline, on the other
	hand, requires up to $367$ CPU cores (i.e., 12 CPU server nodes) to match
	\ours's preprocessing performance, incurring much higher power consumption as well as
	cost.
	\fig{fig:energy_cost_efficiency}(a) summarizes how all of this translate into
	energy consumption. Overall, \ours provides an average $11.3\times$ (maximum
			$15.1\times$) energy-efficiency improvement, demonstrating its merits.

{\bf Cost-efficiency (TCO).}  
\fig{fig:energy_cost_efficiency}(b) compares the cost-efficiency of \ours and \baseline
for data preprocessing (as defined in \sect{sect:method_metric}).  Overall, \ours provides an average $4.3\times$
(maximum $5.6\times$) improvement in cost-efficiency vs. \baseline.
Cost-efficiency is primarily determined by both CapEx and OpEx and our
experiments thus far have demonstrated that \ours outperforms \baseline on
both fronts, providing significant reduction in TCO.

\subsection{PreSto vs. Alternative Accelerated Preprocessing}

\begin{figure}[t]
	\centering
	\includegraphics[width=0.48\textwidth]{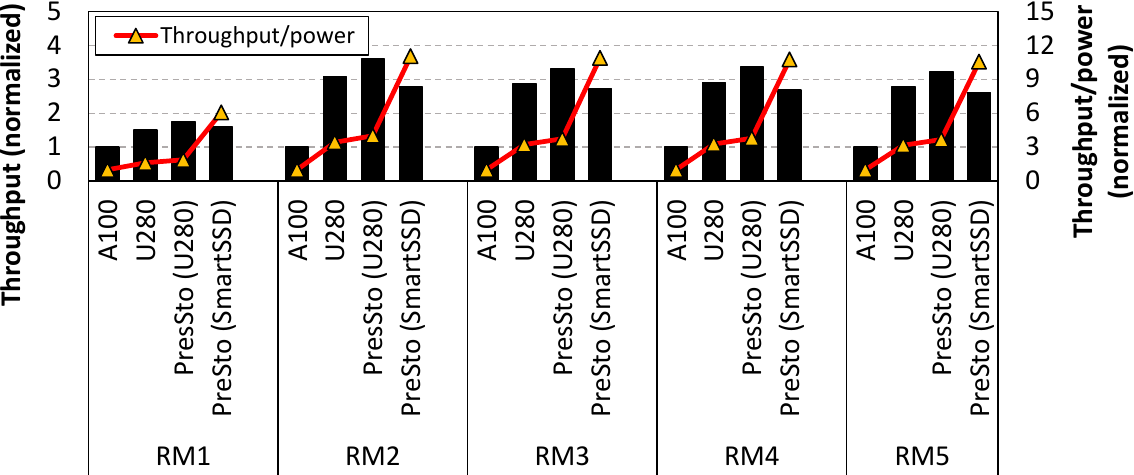} 
	\caption{Data preprocessing's (left axis) performance and (right axis) performance/Watt
		  over \ours (single SmartSSD device),  \ours(single U280 FPGA), a single A100 GPU and a single U280 FPGA.
  }
	\vspace{-1.5em}
	\label{fig:gpu_preprocessing}
  \end{figure}

\begin{figure}[t]
	\centering
	\includegraphics[width=0.48\textwidth]{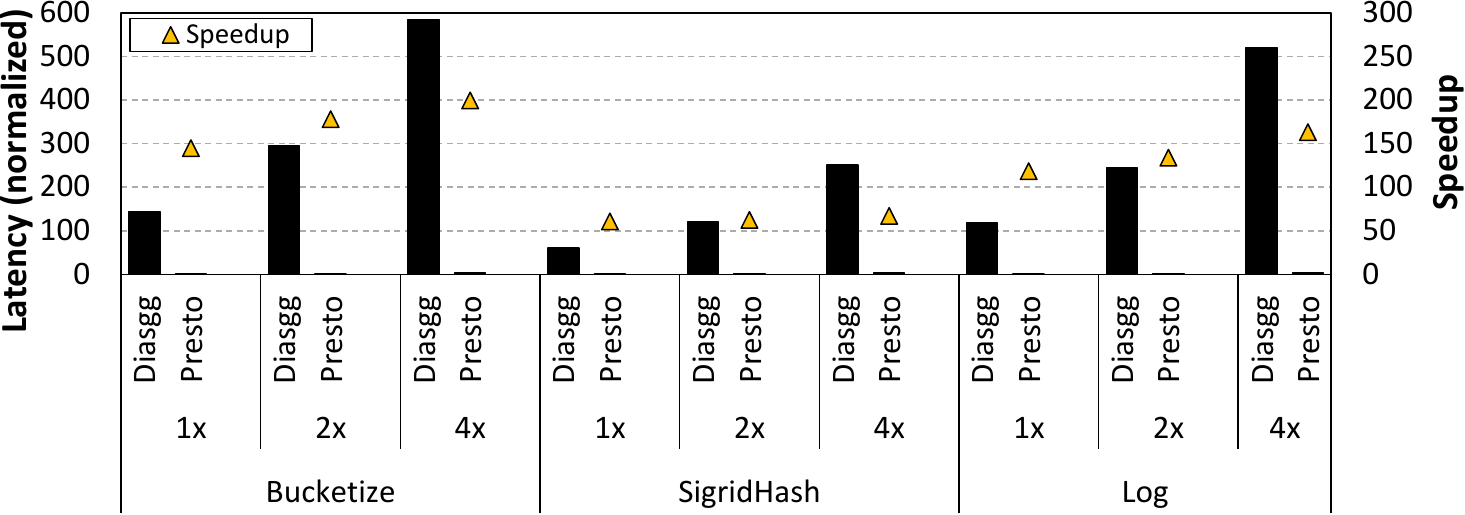} 
	\caption{Latency of \baseline and \ours's feature generation/normalization time (left axis) and \ours's speedup (right axis) when the 
	the number of features for preprocessing is changed. The ``$1\times$'' data point corresponds to the RM5 configuration in \tab{tab:model}. \baseline's latency to conduct each operation is normalized to its respective ``$1\times$'' latency of \ours.
  }
	\vspace{-1.5em}
  \label{fig:sensitivity_num_features}
  \end{figure}
  
Our study thus far have focused on comparing data preprocessing over the
baseline disaggregated CPU servers and ISP architectures.
	For the completeness of our study, we also evaluate the efficacy of alternative accelerated preprocessing solutions where high-end GPUs/FPGAs 
	function as preprocessing accelerators. \fig{fig:gpu_preprocessing} compares the preprocessing performance (left axis)
	and performance/Watt (energy-efficiency) with four system design points: (1) a single A100 GPU (denoted ``A100'') and (2) a single Xilinx U280\cite{u280} FPGA (denoted ``U280'') employed within a disaggregated accelerator pool (\fig{fig:system_designs}(b)), as well as our proposed FPGA-based accelerator system implemented using (3) a single, discrete U280 FPGA card integrated within the storage node over PCIe (denoted ``\ours(U280)'') and (4) one implemented using
a single SmartSSD device (denoted ``\ours(SmartSSD)'').
	We utilizes NVIDIA's NVTabular library~\cite{nvtabular:github} for GPU-based data preprocessing. The U280-based FPGA accelerator is 
	synthesized with $2\times$ larger number of  Decoder, Feature generation, and Feature normalization units that maximally utilize 
	U280's larger custom logics. \ours(SmartSSD) provides an average $2.5\times$ speedup vs. A100 while experiencing an average $5\%$ performance loss vs. U280
	FPGA. Note that \ours(SmartSSD) is able to achieve such performance despite its
	much lower power consumption (TDP of $25$ Watts (SmartSSD) vs. $250$ Watts
	(A100) and 225 Watts (U280)). In general, GPUs are
	throughput-optimized devices with high compute and memory throughput, so they
	perform best when the target application requires massive compute and memory
	accesses. We observe that the compute and memory operations entailed in
	RecSys preprocessing is lightweight vs. training. This makes it
	challenging for the GPU to amortize the cost of CUDA kernel launches, each of
	which has a small working set with modest compute/memory operations, leading
	to significant GPU underutilization. U280 does much better than the GPU
	but its end-to-end speedup vs. \ours(SmartSSD) is relatively low, due to its
	high latency overhead in copying data in/out of the disaggregated preprocessing
	node (which accounts for an average $47.6\%$ of its end-to-end preprocessing time). 
	Even though \ours(U280) minimizes such redundant data movements and achieves a slightly higher preprocessing throughput compared to \ours(SmartSSD),
	\ours(SmartSSD) delivers much higher energy-efficiency (an average $2.9 \times$) vs. \ours(U280) by being custom-designed to right-size its compute units 
	for data preprocessing under a tighter power budget (25 Watts).

\subsection{PreSto Sensitivity to the Number of Features to Preprocess}
\label{sect:sensitivity_num_features}

We investigate \ours's sensitivity to the number of features to preprocess by evaluating the latency incurred in executing the Bucketize, SigridHash, and Log operations
when the number of generated, sparse, and dense features are changed. \fig{fig:sensitivity_num_features} illustrates a comparison of the average latency 
to execute each operation using \baseline and \ours. While the latency of \baseline increases almost proportionally with the number of features for preprocessing, 
 \ours does a much better job leveraging inter-/intra-feature parallelism and consistently achieves significant speedups, demonstrating the robustness of our proposal.

\section{Related work}
\label{sect:related_work}

There is a large body of prior
literature exploring DNN preprocessing, RecSys model training/inference, RecSys data storage and preprocessing, and domain-specific/general-purpose ISP designs, which we summarize below.

{\bf DNN preprocessing.} There are multiple prior work addressing
the performance gap between DNN model training and data
preprocessing~\cite{vldb2021_mohan, tf_data, nvidia_dali, trainbox,
	disaggregation_google, plumber_mlsys2022, cachew, fastflow, quiver,
	dlbooster,icache,preGNN}. Mohan et al.~\cite{vldb2021_mohan} and Murray et
	al.~\cite{tf_data} proposed software optimizations to address this problem,
	while TrainBox~\cite{trainbox} and DALI~\cite{nvidia_dali} proposed hardware
	accelerated preprocessing tackling computer vision and audio training tasks. Similarly, DLBooster~\cite{dlbooster} focused on offloading preprocessing
	operations to an FPGA for inference. PreGNN\cite{preGNN} proposed offloading graph neural network (GNN) preprocessing tasks to an accelerator.
	Several prior	art~\cite{disaggregation_google, plumber_mlsys2022, cachew, fastflow, quiver}
	proposed disaggregated preprocessing solutions but these works primarily
	focused on efficiently managing CPU resources via software optimizations
	using data caching or prefetching. Importantly, all of these prior art
	strictly do not focus on RecSys, rendering the key
	contribution of our work unique.

{\bf RecSys data storage and preprocessing.} Zhao et al.~\cite{dsi} discusses
a disaggregated data preprocessing service for Meta's RecSys training pipeline.
XDL~\cite{xdl} proposed a distributed ML 
framework for Alibaba's production RecSys model.
InTune~\cite{intune_recsys2023} presents a reinforcement learning-based RecSys
data pipeline optimization to efficiently manage CPU resources. There also
exists prior work exploring feature deduplication to improve the performance of
RecSys data preprocessing~\cite{recd}.  While not targeting the preprocessing stage of RecSys,
			 Tectonic-shift~\cite{tectonic_shift} explored the viability of a flash
			 storage tier in the data storage system of Meta's production ML training
			 infrastructure, aiming to improve the performance and power efficiency of
			 their I/O operation in data storage stage. Overall, the contribution of \ours is
			 orthogonal to these studies.

{\bf RecSys model training/inference.}
The surge of interest in RecSys in both academia and industry has spawned numerous prior work accelerating RecSys training and inference utilizing near-/in-memory
processing~\cite{tensordimm,recnmp,tensorcasting,fafnir,trim} as well as various hardware/software optimizations
~\cite{tcaching, merci, deeprecsys, facebook_dlrm, udit:resys:hpca2020:industry_track, 
centaur, acun:resys:hpca2021:industry_track,isca2022_mudigere,hercules,mprec,recpipe,facebook_hpca2018}. 
Importantly, \ours stands apart from this body of work by focusing on accelerating RecSys preprocessing, which is as
discussed in this paper 
completely orthogonal operation compared to model training/inference.

{\bf Domain-specific/general-purpose ISP designs.} There is a large body of prior literature exploring domain-specific/general-purpose ISP designs. GLIST\cite{glist} proposed an ISP architecture for SSD-based GNN inference. SmartSAGE\cite{smartsage} proposed an ISP-based GNN training system to overcome I/O bottleneck of SSD-based training. Mahapatra et al.\cite{dscs_arxiv} proposed an ASIC-based ISP in a disaggregated storage system for serverless functions, alleviating communication overhead of remote storage systems. RecSSD\cite{recssd} and RM-SSD\cite{rm_ssd} proposed ISP to overcome the memory  bottlenecks of RecSys inference. GraphSSD\cite{graphssd} proposed an ISP architecture for graph semantics with a simple programming interface. ECSSD\cite{ecssd} proposed an ISP architecture for extreme classification based on the approximate screening algorithm. Work by Hu et al.\cite{hu2019dynamic} proposed an ISP-based dynamic multi-resolution storage system to mitigate the performance bottleneck of data preparation for approximate compute kernels. ASSASIN\cite{assassin}, INSPIRE\cite{inspire}, and GenStore\cite{genstore} proposed an ISP architecture for stream computing, private information retrieval, and genome sequence analysis, respectively. There also exists a large body of prior literature targeting ISP acceleration for data-intensive workloads. Morpheus\cite{morpheus:isca2016}, DeepStore\cite{deepstore}, Active Flash\cite{active_flash}, GraFBoost\cite{grafboost:isca2018}, Biscuit\cite{biscuit}, and BlueDBM\cite{bluedbm} proposed a domain-specific ISP architecture that targets data analytics, data management, object (de)serialization, or graph analytics. Summarizer\cite{summarizer} and INSIDER\cite{insider} proposed a hardware/software co-designed ISP architecture to offload data-intensive tasks with a set of flexible programming APIs. Willow\cite{willow} proposed architectural support to enhance the effectiveness and flexibility of a programmable ISP design targeting I/O-intensive applications. Unlike these prior work, PreSto demonstrates the merits of an ISP solution targeting \emph{compute-bound} RecSys data preprocessing and uncovers its new system-level bottlenecks, rendering our key contributions unique.

\section{Conclusion}
\label{sect:conclusion}

In this work, we propose an ISP based RecSys data preprocessing system called
\ours which conducts the preprocessing operation close to where the training samples are preserved. By fully leveraging inter-/intra-feature
parallelism available in feature generation/normalization, \ours can
effectively close the performance gap between preprocessing and model training
at a much lower cost and power consumption compared to the baseline CPU-centric
system. Overall, \ours outperforms state-of-the-art preprocessing systems with 
$9.6\times$
speedup in end-to-end preprocessing time, $4.3\times$ improvement in cost-efficiency,
				and $11.3\times$ enhancement in energy-efficiency.

\section*{Acknowledgements}
This work was supported by the National Research Foundation of Korea (NRF) grant funded by the Korea government (MSIT) (NRF-2021R1A2C2091753), and Institute of Information \& communications Technology Planning \& Evaluation (IITP) grant funded by the Korea government(MSIT) (No. 2022-0-01037, Development of High Performance Processing-in-Memory Technology based on DRAM). We also appreciate the support from SNU-SK Hynix Solution Research Center (S3RC) and the EDA tool supported by the IC Design Education Center(IDEC), Korea. Minsoo Rhu is the corresponding author.

\bibliographystyle{IEEEtranS}
\bibliography{refs}

\vspace{12pt}

\end{document}